\documentclass[aps, prb, showpacs, superscriptaddress, twocolumn]{revtex4}
\usepackage{hyperref}
\usepackage[caption=false]{subfig}
\usepackage{graphicx}
\usepackage{amsmath}
\usepackage{amssymb}
\usepackage{amsfonts}   
\usepackage{multirow}

\usepackage{braket}
\usepackage{color}
\renewcommand{\Re}{\operatorname{Re}}

\newcommand{\Tr}{\mathrm{Tr}}

\begin{document}
\title{Hierarchical equations of motion approach to transport through an Anderson impurity coupled to interacting Luttinger liquid leads}

\author{Jun-ichi Okamoto}
\email{ojunichi@physnet.uni-hamburg.de}
\affiliation{Zentrum f\"ur Optische Quantentechnologien and Institut f\"ur Laserphysik, Universit\"at Hamburg, 22761 Hamburg, Germany}
\affiliation{The Hamburg Centre for Ultrafast Imaging, Luruper Chaussee 149, 22761 Hamburg, Germany}

\author{Ludwig Mathey}
\affiliation{Zentrum f\"ur Optische Quantentechnologien and Institut f\"ur Laserphysik, Universit\"at Hamburg, 22761 Hamburg, Germany}
\affiliation{The Hamburg Centre for Ultrafast Imaging, Luruper Chaussee 149, 22761 Hamburg, Germany}

\author{Rainer H\"{a}rtle}
\affiliation{Institut f\"{u}r Theoretische Physik, Georg-August-Universit\"{a}t G\"{o}ttingen, Friedrich-Hund-Platz 1, D-37077 G\"{o}ttingen, Germany}

\date{\today}

\begin{abstract}
We generalize the hierarchical equations of motion method to study electron transport through a quantum dot or molecule coupled to one-dimensional interacting leads that can be described as Luttinger liquids. Such leads can be realized, for example, by quantum wires or fractional quantum Hall edge states. In comparison to noninteracting metallic leads, Luttinger liquid leads involve many-body correlations and the single-particle tunneling density of states shows a power-law singularity at the chemical potential. Using the generalized hierarchical equations of motion method, we assess the importance of the singularity and the next-to-leading order many-body correlations. To this end, we compare numerically converged results with second and first-order results of the hybridization expansion that is inherent to our method. As a test case, we study transport through a single-level quantum dot or molecule that can be described by an Anderson impurity model. Cotunneling effects turn out to be most pronounced for attractive interactions in the leads or repulsive ones if an excitonic coupling between the dot and the leads is realized. We also find that an interaction-induced negative differential conductance near the Coulomb blockade thresholds is slightly suppressed as compared to a first-order and/or rate equation result. Moreover, we find that the two-particle ($n$-particle) correlations enter as a second-order ($n$-order) effect and are, thus, not very pronounced at the high temperatures and parameters that we consider. 
\end{abstract}

\maketitle
\section{Introduction}
\label{intro}
Quantum dots and molecules offer unique opportunities to design nanoelectronic devices and to understand fundamental properties of open quantum many-body systems such as quantization, interference, 
exchange and nonequilibrium effects.\cite{Kastner2000,Aleiner2002,Fujisawa2006,cuevasscheer2010,Baldea15} Conventional studies 
focus on the coupling to noninteracting leads, which is appropriate when the electrons in the leads are Fermi liquids. \cite{datta1997electronic,Andergassen2010} 
In contrast, interacting electrons in a quasi-one-dimensional lead can organize themselves into a Luttinger liquid giving rise to a unique set of correlations and properties. \cite{Haldane1981,giamarchi2003quantum} The importance of such correlations on the corresponding transport properties can be demonstrated by Kane-Fischer theory, \cite{Kane1992, Kane1992a} which shows that the conductance through a tunneling barrier vanishes if it is attached to repulsive Luttinger liquid leads, yet becomes perfect for attractive ones. Experimentally, it is challenging to realize Luttinger liquid leads, while the interest in such systems is growing due to possible realizations in quantum wires, \cite{Auslaender2000,Matzdorf2009,Matzdorf2011,Laroche2014} or by the edge states of a fractional quantum Hall system. \cite{Milliken1994,Chang2003,Wurstbauer2013,Pascher2014,Li2015} Thus, it is very interesting to study the (non-equilibrium) properties of such transport setups.

Transport with Luttinger liquid leads has been investigated before. \cite{vonDelft1998,giamarchi2003quantum,Schoenhammer2005,Metzner2012} Two Luttinger liquids separated by a tunneling barrier were first studied by Kane and Fisher,\cite{Kane1992, Kane1992a} and later by others. \cite{Furusaki1993, Matveev1993, Yue1994, Moon1995, Egger1995, Fendley1995, Furusaki1997, Aristov2008, Aristov2009, Einhellinger2012} Fabrizio et al. \cite{Fabrizio1994} and Maurey and Giamarchi \cite{Maurey1995} extended these studies by long-range Coulomb interactions in the leads. noninteracting structures or quantum dots with a single-level that replaces the tunneling barrier of the latter studies were also considered,\cite{Weiss1995, Nazarov2003, Komnik2003, Komnik2003a, Waechter2007, Goldstein2010, Goldstein2010b} including the dual, side-coupled case. \cite{Lerner2008,Goldstein2010a} Here, the position of the energy level is crucial; a level aligned with the chemical potential yields perfect conductance while the conductance is zero otherwise. 
Transport with a single level quantum dot, which can be described by an Anderson impurity model has 
been investigated by Andergassen et al.,\ \cite{Andergassen2006} where the authors find that screening of the dot spin (Kondo physics) occurs even in the presence of the power-law singularity inherent to a Luttinger liquid.\cite{Haldane1981,Jeckelmann2012} 
These considerations were extended to Coulombic dot-lead interactions by Elste et al.,\cite{Elste2011a} 
revealing a mechanism for negative differential resistance (NDR) at the Coulomb blockade edge (similar to the NDR mechanisms reported by Matveev and Larkin \cite{Matveev1992} and Dubi \cite{Dubi2013}). 
In this work, we study the same model; we consider transport through an Anderson impurity coupled to two Luttinger liquid leads 
where all interactions are local and short-ranged. Note that transport through more complex, multi-orbital systems has also been considered,\cite{Furusaki1998,Durganandini2006,Durganandini2006b,Kawaguchi2009,Yang2014} including the effect of electron-phonon interactions.\cite{Maier2010,Yang2010,Skorobagatko2012} 

The early studies on the low-energy transport properties of a tunneling barrier connected to Luttinger liquid leads are 
based on analytic results. They have been derived either by employing Bethe ansatz solution \cite{Fendley1995} or by 
bosonic \cite{Kane1992,Furusaki1993} or fermionic renormalization group.\cite{Matveev1993, Yue1994, Aristov2008, Aristov2009}
Approximate solutions on the more complex setups with double barrier or quantum dot structures have been derived using rate equations,\cite{Furusaki1998,Durganandini2006,Durganandini2006b,Elste2011a} functional renormalization group,\cite{Andergassen2006,Waechter2007} equation of motion techniques,\cite{Skorobagatko2012} full counting statistics \cite{Maier2010} or 
nonequilibrium Green's functions.\cite{Kawaguchi2009,Yang2010,Yang2014} Numerically converged or exact results have become available only recently. 
This includes a study of the tunneling barrier case by density matrix renormalization group (DMRG) \cite{Einhellinger2012} and 
an imaginary-time quantum Monte Carlo (QMC) scheme where a quantum dot is described that is coupled to a single Luttinger liquid lead. \cite{Hattori2014} 
Thus, the number of numerically exact schemes that can be used to describe this transport setup is very limited. This is 
in contrast to the transport setups with noninteracting leads, where density matrix renormalization group,\cite{Hassanieh2006,daSilva2008,Kirino2008,HeidrichMeisner2009,Kirino2010,Branschadel2010} 
numerical renormalization group, \cite{Anders2008,Anders2008b,Schmitt2010,Nghiem2014} 
multilayer multiconfiguration time-dependent Hartree theory, \cite{Thoss2013,Balzer2015} iterative \cite{Thorwart2008,Segal2010,Segal2011,Huetzen2012,Weiss2013} and stochastic path-integral 
schemes \cite{Werner2006,Schmidt2008,Werner2009,Schiro2010,Gull2010,Cohen2011,Muhlbacher2011,Cohen2013} or 
the hierarchical quantum master equation (HQME) method \cite{Jin2008,Zheng2009,Hartle2013,Hartle2015,Cheng2015,Wenderoth2016} have been employed.

In this paper, we extend the HQME framework to describe transport with Luttinger liquid leads. 
The method was originally developed to describe finite quantum systems coupled to a bosonic 
environment \cite{Tanimura1989,Tanimura1990,Tanimura2006} and was later extended to fermionic 
reservoirs. \cite{Welack2006,Jin2008,Hartle2013,Hartle2015} 
The method is based on a hybridization expansion that, for noninteracting leads, can be systematically 
converged if the temperature of the environment is not too low (for a single-impurity Anderson model, the numerical effort becomes prohibitive below the Kondo temperature). \cite{Hartle2015} 
Thus, higher order processes between the dot and the leads, and also correlations 
in the dot and the leads, can be accounted for systematically. In parallel, a perturbative analysis to a fixed, given order 
is also possible. We present a scheme that allows to obtain converged results with respect to single-particle correlations 
in the Luttinger liquid leads, including the effect of the power-law singularity. The effect of two- and $n$-particle correlations, 
however, is included only in leading order. Thus, our method represents an important step towards a numerically exact scheme and 
allows us to estimate, at least, the role of multi-particle correlations. 
In contrast to DMRG or imaginary-time QMC, it also allows us to account for the full nonequilibrium character of this transport problem. 
As a standard example, we study the transport properties of a single-level quantum dot 
or molecule that can be described by the Anderson impurity model. For this model, we corroborate previous results on NDR at the Coulomb edge, \cite{Elste2011a} assessing the role of cotunneling and beyond-second-order processes and 
the role of two-particle correlations in the leads. Processes beyond second order and two-particle correlations turn out to be 
negligible in the range of temperatures that we consider ($T\gg T_\text{Kondo}$).

The rest of the paper is organized as follows. In Sec.~\ref{model}, the model that we study is introduced. In Sec.~\ref{method}, we outline the derivation of the HQME with particular emphasis on the differences between the hierarchy construction for noninteracting and interacting leads. 
We also compare our new method with the rate equation scheme of Elste et al.. \cite{Elste2011,Elste2011a} We present our results in Sec.~\ref{results}, where we discuss current-voltage characteristics in the steady state for various interaction types and strengths in the leads. The effects of two-particle correlations functions, which become important when interactions in the leads are non-negligible, are also discussed. We present 
our conclusions in Sec.~\ref{conclusion}. Technical details are outlined in the Appendices.

\section{Model}
\label{model}
\begin{figure}[tb]
  \includegraphics[width=\columnwidth]{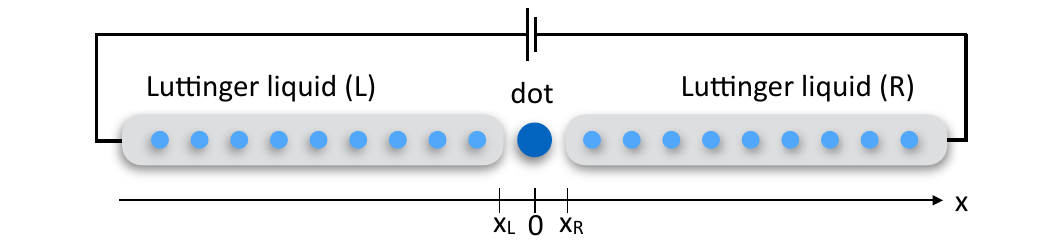}
\caption{Graphical scheme of the transport setup that we consider, i.e., a quantum dot coupled to two, semi-infinite Luttinger liquid leads with chemical potentials that differ by the applied bias voltage. }
\label{fig:model}
\end{figure}
The model that we consider (see Fig.~\ref{fig:model}) is described by the Hamiltonian 
\begin{equation}
H  = H_\text{dot} +  \sum_{\alpha = L, R} \left( H_\text{lead}^{\alpha} + H_\text{tun}^{\alpha} \right).
\end{equation}
It includes a quantum dot (located at position $x=0$) that can be described by an Anderson impurity model with a spin-degenerate level $\epsilon$ and an on-dot Coulomb repulsion $U$,
\begin{equation}
H_\text{dot} = \epsilon \sum_{s = \uparrow, \downarrow} n_s  +U n_{\downarrow} n_{\uparrow},
\end{equation}
where $n_{s} = d^{\dagger}_s d_s$ is the electron density with spin $s$, and $d^{(\dagger)}_s$ is the dot annihilation (creation) operator with spin $s$. 

The dot is coupled to two, semi-infinite Luttinger liquid leads that terminate at $x_{\alpha = L, R}$. We model the coupling between the dot and the leads by a single-particle tunneling Hamiltonian 
\begin{equation}
H_\text{tun}^{\alpha} = \sum_{ s} \int_{\alpha} dx \left[ \mathcal{T}_{\alpha} (x)  \psi_{\alpha s}^{\dagger} (x) d_{s} +\text{H.c.} \right],
\end{equation}
where $\mathcal{T}_{\alpha } (x)$ is the tunneling amplitude and $\psi_{\alpha s}^{(\dagger)} (x)$ is the electron annihilation (creation) operator in the lead $\alpha$. The integral $\int_{\alpha}$ covers the range from $-\infty$ to $x_L <0$ for the left lead, and from $x_R >0$ to $\infty$ for the right lead. This field-theoretical form of the tunneling Hamiltonian is often used in the description of Luttinger liquid leads. It reduces to a simpler form  
\begin{equation}
H_\text{tun}^{\alpha} (t) =  \sum_{s} f^{+}_{\alpha  s } (t)d_{ s} + \text{H.c.},
\end{equation}
where $f_{\alpha s}^{\sigma}$ is a creation ($\sigma = +$) and annihilation ($\sigma = -$) operator in the interaction picture defined as
\begin{equation}
\begin{split}
f_{\alpha s}^{-(+)} (t) &\equiv e^{i H_\text{lead}^{\alpha} t} \left( \int_{\alpha} dx  \mathcal{T}_{\alpha} (x)  \psi_{\alpha s}^{(\dagger)} (x) \right) e^{-i H_\text{lead}^{\alpha} t} \\
&\simeq a \bar{\mathcal{T}}_{\alpha}   \bar{\psi}_{\alpha s}^{(\dagger)} (t) ,
\end{split}
\end{equation}
if one considers tunneling amplitudes that are non-zero only within a small range $a$ at the edge of each lead. $\bar{\mathcal{T}}_{\alpha}$ and $\bar{\psi}_{\alpha s}^{(\dagger)} (t)$ represent the average tunneling amplitude and field strength over this small range, respectively.

The leads are, for example, Hubbard chains whose low-energy properties can be described by an effective model of a Luttinger liquid. We do not assume a specific form of a microscopic Hamiltonian, because only the effective Luttinger form is used in the following. We set the notation using the bosonized Hamiltonian for an infinite lead \cite{giamarchi2003quantum}
\begin{equation}
H_\text{lead}^{\alpha} =   \sum_{\nu = c, s}  \frac{1}{2\pi} \int  dx \left[ u^{\alpha}_{\nu} K^{\alpha}_{\nu} \left( \nabla \theta^{\alpha}_{\nu}\right)^2 + \frac{u^{\alpha}_{\nu}}{K^{\alpha}_{\nu}} \left( \nabla \phi^{\alpha}_{\nu}\right)^2 \right] ,
\end{equation}
where $\phi_{s,c}$ and $\theta_{s,c}$ are conjugate bosonic variables for spin and charge sectors,  $u^{\alpha}_{\nu}$ is the renormalized velocity of sound, and $K^{\alpha}_{\nu}$ are Luttinger parameters. For a Hubbard chain, for example, the Luttinger parameter $K^{\alpha}_c$ is given by the on-site interaction $U_\alpha$ and the bare Fermi velocity $v_{F,\alpha}$ as
\begin{equation}
(K^{\alpha}_c)^2 = \frac{1}{1+U_\alpha/\pi v_{F,\alpha}}.
\end{equation}
We assume a SU(2) invariant system about spins, i.e.\ we set the effective Luttinger constant $K^{\alpha}_s = 1$ 
and $u^{\alpha}_s$ is equal to $v_{F,\alpha}$. The transformation formula from fermions to bosons is
\begin{equation}
\begin{split}
\psi_{\alpha s } (x,t) &= \sum_{r = \pm} \psi_{\alpha  r s} (x,t) \\
&= \sum_{r = \pm} \eta_{\alpha rs} e^{i \mu_{\alpha} t} \frac{e^{ i r k_{F,\alpha} x}}{\sqrt{2\pi a}} e^{-\frac{i}{\sqrt{2}} \left[ \left( r \phi^{\alpha}_{ c} - \theta^{\alpha}_{ c} \right) + s   \left( r \phi^{\alpha}_{s} - \theta^{\alpha}_{s} \right) \right]},
\end{split}
\end{equation}
where $r$ is the chirality of the fermion, and $\eta_{\alpha r s}$ is a Klein factor taking care of the anticommutation relations among fermions. $\mu_{\alpha}$ denotes the chemical potential of lead $\alpha$.

An essential ingredient for the HQME is the correlation function of the lead electrons at the dot-lead boundary, which is calculated over the equilibrium density matrix of the leads $\rho_{L+R}$ as $\braket{\cdot}_{L+R} \equiv \Tr_{L+R} \left[ \rho_{L+R} \cdots \right]$. Including the boundary condition for the wave functions $\psi_{L s}^{(\dagger)} (x >x_L) = \psi_{R s}^{(\dagger)} (x < x_R) = 0$, the correlation functions at the dot-lead boundary can be calculated as \cite{Furusaki1998, Elste2011, Elste2011a} (from now on we drop the spatial dependence in the field operators)
\begin{equation}
\begin{split}
C^{\sigma}_{\alpha  } (t) &\equiv \left\langle f_{\alpha s}^{\sigma} (t) f_{\alpha  s}^{\bar{\sigma}} (0)  \right\rangle_{L+R} \\
&= \Gamma_{\alpha} e^{i \sigma \mu_{\alpha} t} \left\{ \frac{i \hbar v_{F,\alpha}  \beta_{\alpha}}{\pi a} \sinh \left[ \frac{\pi (t - i \delta)}{\beta_{\alpha} \hbar} \right] \right\}^{-Y_{\alpha}},
\end{split}
\label{psi_corr}
\end{equation}
where $\Gamma_{\alpha} = a |\bar{\mathcal{T}}_{\alpha}|^2/ \pi$ is the system-lead coupling and $\bar{\sigma} \equiv - \sigma$. $\delta$ is a short-time cut-off, which we choose $\delta = 1/W_{\alpha}$ with the band width $W_{\alpha}$ that determines the scale where the fermionic dispersion can be considered linear, $W_{\alpha} \simeq v_{F,\alpha} /a $. $\beta_{\alpha}$ is the inverse temperature of lead $\alpha$. The parameter $Y_{\alpha} = 1/(2K^{\alpha}_c)+1/2$ determines the character and strength of the interactions in the Luttinger liquid in the charge sector. $Y_{\alpha}$ greater (smaller) than 1 represents repulsive (attractive) interactions in the Luttinger liquid. Elste et al.\ \cite{Elste2010, Elste2011a} 
showed that a short-range excitonic (density-density) coupling between the dot and the leads effectively renormalizes $Y_{\alpha}$ to
\begin{equation}
Y_{\alpha} = \frac{\left(1 - \frac{K^{\alpha}_c}{u^{\alpha}_c} V_{\alpha} \right)^{2}}{2 K^{\alpha}_c} + \frac{K^{\alpha}_c V_{\bar{\alpha}}^2 }{2 u_c^{\alpha,2}} + \frac{1}{2},
\label{Y_renorm}
\end{equation}
where $V_{\alpha}$ is the excitonic coupling strength. This means that an originally repulsive lead, which is usually the case for quantum wires and carbon nanotubes, can be effectively attractive. Thus in the following we consider both repulsive and attractive cases. The excitonic coupling also renormalizes the energy $\epsilon$ and the interaction strength $U$, but we assume that such effects are already included in these parameters. 

\begin{figure}[tb]
  \includegraphics[width=0.9 \columnwidth]{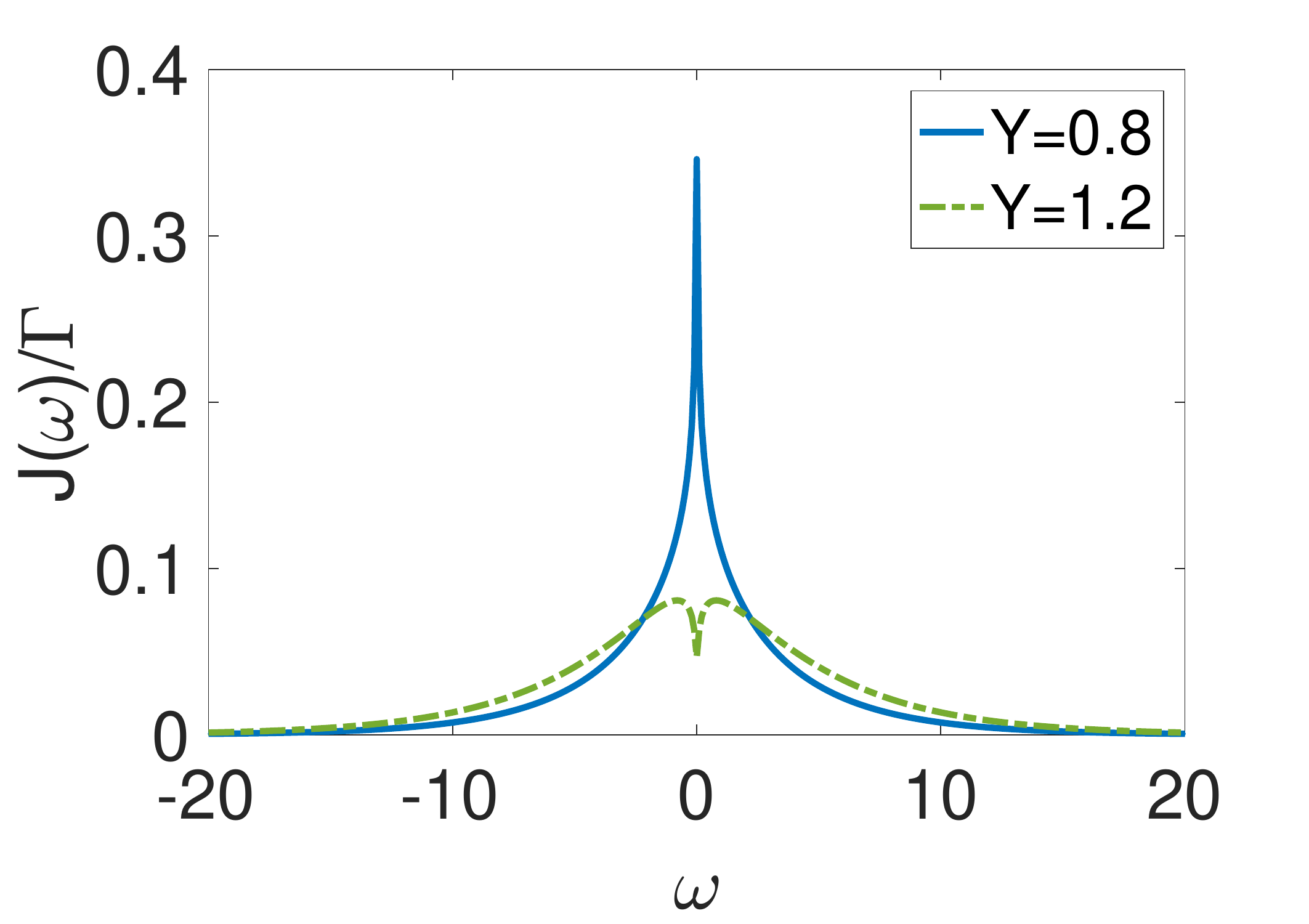}
\caption{Single-particle tunneling density of states, $J(\omega)$, at the edge of a Luttinger liquid lead for attractive ($Y=0.8$) and repulsive interactions ($Y=1.2$) with $T = 200$\,K, and $W = 4$\,eV. $\Gamma$ is the system-lead coupling strength.}
\label{fig:psd}
\end{figure}

We note that the correlation function, Eq.~\eqref{psi_corr}, gives a power-law singularity in the single-particle tunneling density of states at low frequencies and low temperatures,
\begin{equation}
\begin{split}
J_{\alpha}(\omega) &\equiv \int_{-\infty}^{\infty} \frac{dt}{2\pi} e^{-i \omega t} C^{+}_{\alpha}(t) + \text{h.c.} \\
&\propto |\omega - \mu_{\alpha}|^{Y-1}  \ \ \text{as} \ \ |\omega - \mu_{\alpha}| \ll W_{\alpha}, T_{\alpha}.
\end{split}
\end{equation}
Numerical examples for the single-particle tunneling density of states are depicted in Fig.~\ref{fig:psd}. While attractive interactions ($Y=0.8$) induces a sharp peak at the chemical potential, $\omega =0$, repulsive interactions ($Y=1.2$) lead to a sharp dip. This dependence of the low energy single-particle tunneling density of states on the interaction parameter $Y_{\alpha}$ crucially affects the transport properties, even at the single-particle level.

\section{Method}
\label{method}
\subsection{Hierarchical equations of motion}
In this section, we briefly outline the derivation of the hierarchical equations of motion for transport with Luttinger liquid leads. The derivation is quite general, and not necessarily restricted to Luttinger liquids, while, to implement the formalism in practice, we make use of the specific properties of Luttinger liquids. Note that our derivation follows closely Refs.\ \onlinecite{Jin2008} and \onlinecite{Hartle2013}. 

We first trace out the lead variables from the total density matrix $\rho_T (t)$ of the system (the dot and leads) to obtain an equation of motion for the reduced density matrix for the dot degrees of freedom,
\begin{equation}
\rho (t) \equiv \Tr_{L+R} \left [ \rho_T (t)\right] .
\end{equation}
Formally the time propagation of the reduced density matrix can be written as $\rho (t) = \mathcal{U} (t, t_0) \rho (t_0)$, where $\mathcal{U} (t, t_0)$ can be written in terms of a path integral representation as 
\begin{equation}
 \mathcal{U} (t, t_0) = \int_{\xi_0}^{\xi} \mathcal{D} \tilde{\xi} \int_{\xi'_0}^{\xi'} \mathcal{D}  \tilde{\xi}' e^{ i S[\tilde{\xi}]} \mathcal{F}[\tilde{\xi}, \tilde{\xi}']e^{- i S[\tilde{\xi}']},
\end{equation}
where $\ket{\xi} \equiv \exp \left(\sum_{ s} \xi_{s} d^{\dagger}_{ s} \right) | 0 \rangle$ is a fermionic coherent state, and $S$ is the action of the dot Hamiltonian. $\mathcal{F} $ is the Feynman-Vernon influence functional
\begin{multline}
\mathcal{F} = \Bigg\langle T \exp \left[ -i \int_{t_0}^t d\tau  \sum_{\alpha,  s} f^{+}_{\alpha  s } (\tau) \tilde{\xi}_{s}(\tau) + \text{H.c.}   \right]   \\   \times \bar{T} \exp \left[ i \int_{t_0}^t d\tau  \sum_{\alpha,   s} f^{+}_{\alpha   s } (\tau) \tilde{\xi}_{s}(\tau) + \text{H.c.}   \right] \Bigg\rangle_{L+R} ,
\label{FV}
\end{multline}
with time-ordering $T$ and anti-time-ordering $\bar{T}$. Expanding the exponentials in Eq.~\eqref{FV} and integrating the lead variables, $\mathcal{F}$ contains a series of correlation functions of $f^{(\dagger)}_{\alpha s}$. Let us schematically write this as
\begin{equation}
\mathcal{F} \sim 1 + \int C^{(1)} \tilde{\xi}^2 + \int C^{(2)}\tilde{\xi}^4 + \dots,
\label{FV_exp}
\end{equation}
where $C^{(k)}$ denotes a $k$-particle correlation function (including disconnected diagrams). The formally exact series expansion of $\mathcal{F}$ is given in Appendix \ref{appA}. We can reorganize the expansion by using the influence exponent 
\begin{equation}
\Phi \equiv - \log \mathcal{F}.
\label{replica_id}
\end{equation}
The corresponding expansion of $\Phi$, i.e., the cumulant expansion of $\mathcal{F}$ is found to be a series of connected correlation functions,
\begin{equation}
\begin{split}
\Phi &\sim - \int C^{(1)} \tilde{\xi}^2 - \int C^{(2)}\tilde{\xi}^4 +  \int [C^{(1)}]^2 \tilde{\xi}^4 + \dots \\
&\sim  - \int C^{(1)} \tilde{\xi}^2 - \int \hat{C}^{(2)}\tilde{\xi}^4 + \dots,
\end{split}
\end{equation}
where $\hat{C}^{(2)}$ is the connected two particle correlation functions in the leads. For noninteracting leads, only the first term remains since all the higher order correlation functions can be written as products of $C^{(1)}$ by Wick's theorem, and are summed up to be zero. In the Appendix\ \ref{appB}, we show an explicit formula for $\Phi$ up to $\hat{C}^{(2)}$ for interacting leads. 
Now using Eq.~\eqref{replica_id}, we can express $\mathcal{F}$ as the exponential of connected correlation functions.

Neglecting ($k>2$)-particle correlations, the equation of motion of $\mathcal{F}$ can thus be written as 
\begin{equation}
\begin{split}
\dot{\mathcal{F}} &\simeq -i \sum_{\sigma \alpha  s} \mathcal{A}^{\bar{\sigma}}_{ s}  \left( \mathcal{B}^{\sigma}_{\alpha s} + \hat{\mathcal{B}}^{\sigma}_{1, \alpha s} + \hat{\mathcal{B}}^{\sigma}_{2, \alpha s} \right) \mathcal{F} \\
&\equiv -i \sum_{\sigma \alpha  s} \mathcal{A}^{\bar{\sigma}}_{ s} \left( \mathcal{F}^{\sigma}_{\alpha s} +  \hat{\mathcal{F}}^{\sigma}_{1, \alpha s} +  \hat{\mathcal{F}}^{\sigma}_{2, \alpha s} \right),
\end{split}
\label{FV_EOM_0}
\end{equation}
with  
\begin{equation}
\begin{split}
&\mathcal{A}^{\sigma}_{s} \equiv  \xi_{ s}^{\sigma}  + \xi_{ s}^{\prime \sigma}, \\
&\mathcal{B}^{\sigma}_{\alpha s}  \equiv -i  \int_{t_0}^{t} d\tau \left[ C^{\sigma}_{\alpha  } (t, \tau) \tilde{\xi}_{s}^{\sigma} (\tau) -  C^{\bar{\sigma} *}_{\alpha  } (t, \tau) \tilde{\xi}_{ s}^{\prime \sigma} (\tau) \right] ,\\
&C^{\sigma}_{\alpha  } (t-t') \equiv \left\langle f_{\alpha s}^{\sigma} (t) f_{\alpha  s}^{\bar{\sigma}} (t')  \right\rangle_{L+R}.
\end{split}
\label{ABC}
\end{equation}
The first term is related to the $C^{(1)}$ correlator in Eq.~\eqref{FV_exp}. The second and third terms are associated with the $\hat{C}^{(2)}$ correlator. Explicit expressions of the $\hat{\mathcal{B}}_{1, \alpha s}^{\sigma}$ and $\hat{\mathcal{B}}_{2,\alpha  s}^{\sigma}$ operators are given in Appendix \ref{appB}. They include the correlation function $\hat{C}^{(2)}$ and three $\tilde{\xi}^{(\prime)}$ Grassmann numbers convoluted in three time integrals. 
In the second line, we define the $1$st tier auxiliary influence functionals (AIF's) $\mathcal{F}^{\sigma}_{\alpha s}$, $\hat{\mathcal{F}}^{\sigma}_{1, \alpha s} $, and $\hat{\mathcal{F}}^{\sigma}_{2, \alpha s} $. They enter the equation of motion of the reduced density matrix by the following set of auxiliary density operators (ADO's) 
\begin{equation}
\begin{split} 
\rho^{\sigma}_{\alpha s} &= \mathcal{U}^{\sigma}_{\alpha s} (t, t_0) \rho(t_0), \\
\mathcal{U}^{\sigma}_{\alpha s} (t, t_0) &= \int_{\xi_0}^{\xi} \mathcal{D} \tilde{\xi} \int_{\xi'_0}^{\xi'} \mathcal{D}  \tilde{\xi}' e^{ i S[\tilde{\xi}]} \mathcal{F}^{\sigma}_{\alpha s}[\tilde{\xi}, \tilde{\xi}']e^{- i S[\tilde{\xi}']},
\end{split}
\end{equation}
with similar relations for $\hat{\rho}^{\sigma}_{1(2), \alpha s}$ and $\hat{\mathcal{F}}^{\sigma}_{1(2), \alpha s}$. The equation of motion of the reduced density matrix in terms of ADO's can thus be written as 
\begin{equation}
\dot{\rho}(t) = -i\mathcal{L} \rho (t) - i \sum_{\alpha  s \sigma} \left[ d^{\bar{\sigma}}_{s},  \rho^{\sigma}_{\alpha s} (t) + \hat{\rho}^{\sigma}_{1, \alpha s} (t) + \hat{\rho}^{\sigma}_{2, \alpha s} (t)  \right] .
\end{equation}
The Louiville term $\mathcal{L} \rho = [H_\text{dot}, \rho ]$ arises from the time derivative of the action $S$ in Eq.~\eqref{FV}. 

Taking the time derivative of ADO's opens an infinite hierarchy of equations of motion, including higher order ADO's. In particular, a new type of ADO's emerges from the time derivative of the correlation functions, for example, $C^{\sigma}_{\alpha  }$ as
\begin{equation}
\tilde{\mathcal{B}}^{\sigma}_{\alpha s}  = -i  \int_{t_0}^{t} d\tau \left[ \dot{C}^{\sigma}_{\alpha  } (t, \tau) \tilde{\xi}_{s}^{\sigma} (\tau) -  \dot{C}^{\bar{\sigma} *}_{\alpha  } (t, \tau) \tilde{\xi}_{ s}^{\prime \sigma} (\tau) \right]\neq  \mathcal{B}^{\sigma}_{\alpha s}
\end{equation}
In order to obtain a closed set of ADO's, the correlation functions $C_{\alpha }^{\sigma} (t )$ are translated to a sum of exponentials (Meier-Tannor parametrization method),
\begin{equation} \begin{split}
C_{\alpha }^{\sigma} (t ) &\equiv \sum_{l = 0}^{\infty} h_{\alpha l } e^{- \omega_{\alpha l }^{\sigma} t},   \\
h_{\alpha  l}&=  |\Gamma_{\alpha}| 2^{Y_{\alpha}} \frac{(Y_{\alpha})_l}{l!}   \left( \frac{W_{\alpha}  \beta_{\alpha}}{\pi} \right)^{-Y_{\alpha}} e^{-i Y_{\alpha} \frac{\pi}{2} } e^{ i (Y_{\alpha} + 2 l )\frac{\pi}{\beta_{\alpha} W_{\alpha}}},  \\
 \omega_{\alpha l }^{\sigma} &=  (Y_{\alpha} + 2 l) \frac{\pi}{\beta } - i \sigma \mu_{\alpha},
\end{split}
\label{C_approx}
\end{equation} 
where $(x)_n \equiv \prod_{k=0}^{n-1} (x+k)$ is the Pochhammer symbol. Correspondingly, $ \mathcal{B}^{\sigma}_{\alpha s}$ is also decomposed into a sum as $ \mathcal{B}^{\sigma}_{\alpha s} = \sum_{l=0}^{\infty}  \mathcal{B}^{\sigma}_{\alpha  l s}$, and 
\begin{equation}
 \mathcal{B}^{\sigma}_{\alpha  l s} =  -i h_{\alpha l } \int_{t_0}^{t} d\tau  e^{- \omega_{\alpha l }^{\sigma} (t-\tau)}\left[  \tilde{\xi}_{s}^{\sigma} (\tau) -  \tilde{\xi}_{ s}^{\prime \sigma} (\tau) \right] .
\end{equation}
Now the zeroth tier equation becomes 
\begin{equation}
\dot{\rho} = -i\mathcal{L} \rho  - i \sum_{j} \left[ d^{\bar{\sigma}}_{s},  \rho_j  + \hat{\rho}_{1, j} + \hat{\rho}_{2,j} \right],
\label{0th_tier}
\end{equation}
where we introduced abbreviated notations $j = \{ \alpha l s \sigma \}$, and $\bar{j} = \{ \alpha l s \bar{\sigma} \}$. Note that it is in general impossible to decompose $\hat{C}^{(2)}$ as in Eq.~\eqref{C_approx}. Therefore, we use a local time approximation to reduce its complexity to an exponential series. This approximation is motivated by the fact that the $\hat{C}^{(2)} (t_1, t_2, t_3, t_4)$ are exponentially small when any two of the four time variables differ by $|t_i - t_j| \geq 1/W$ (see Ref.~\onlinecite{Jang2002} and Appendix \ref{appC} for details); the approximation is thus better in the wide band limit. We now discuss the equations of motion of the two classes of ADO's, $\rho_j$ and $\hat{\rho}_{1(2),j}$, separately. 

We start with the equation of motion of the $\rho_j$ operators, 
ignoring the contributions from ($k>1$)-particle correlations (which is an exact procedure for noninteracting leads). 
It leads to a hiearchy of general $n$th tier ADOs, $\rho_{\mathbf{j}}$, associated with an AIF $\mathcal{B}_{j_n}\mathcal{B}_{j_{n-1}} \cdots \mathcal{B}_{j_1} \mathcal{F}$, and obey the following equations of motion 
\begin{multline}
\dot{\rho}^{(n)}_{\mathbf{j}} = - i \mathcal{L} \rho^{(n)}_{\mathbf{j}} - \left( \sum_{k=1}^n \omega_{j_k} \right)\rho^{(n)}_{\mathbf{j}} \\
- i \sum_{k = 1}^n (-1)^{n-k}  \tilde{\mathcal{C}}_{j_k} \rho^{(n-1)}_{\mathbf{j}_k}- i \sum_{j \not \in \mathbf{j}} \mathcal{A}_{\bar{j}} \rho^{(n+1)}_{\mathbf{j}j}
\label{HQME}
\end{multline}
with superoperators $\tilde{\mathcal{C}}$ and $\mathcal{A}$ as
\begin{equation} \begin{split}
 \tilde{\mathcal{C}}_{j} \rho^{(n)} &= \sum_{\nu}  \left[ h_{\alpha l}  d_{s}^{\sigma}  \rho^{(n)} - (-1)^{n} h^{ *}_{\alpha  l }   \rho^{(n)} d_{ s}^{\sigma}   \right], \\
\mathcal{A}_{\bar{j}} \rho^{(n)} &= d_{ s}^{\bar{\sigma}} \rho^{(n)} +(-1)^{n} \rho^{(n)} d^{\bar{\sigma}}_{ s} .
\end{split}\end{equation} 
The second and third terms originate from the time derivative of exponential time dependence and of the integral limit in $\mathcal{B}_{j_k}$ respectively. In the third term, $\rho^{(n-1)}_{\mathbf{j}_k}$ is an ADO where $\mathcal{B}_{j_k}$ is removed from $\rho^{(n)}_{\mathbf{j}}$. The last term is associated with the time derivative of $\mathcal{F}$, which adds $\mathcal{B}_{j}$ on the left of $\mathcal{B}_{j_n}\mathcal{B}_{j_{n-1}} \cdots \mathcal{B}_{j_1} \mathcal{F}$. The constraint on the summation, $ j \not \in \mathbf{j}$, is due to the anticommutation property of Grassmann numbers. We note that the current is related to the first tier ADO as \cite{Jin2008}
\begin{equation}
I_{\alpha} = i \sum_{l, s} \Tr{\left[ \left( \rho_{\alpha l s}^{\dagger}  + \hat{\rho}_{1, \alpha l s}^{\dagger}  + \hat{\rho}_{2, \alpha l s}^{\dagger}  \right)d_{s} - \text{h.c.} \right] }.
\end{equation}
The contributions from the operators $\hat{\rho}$ always conserves the current, i.e., $I_L = - I_R$, which can be easily seen from the equation of motions of $\hat{\rho}$ considered next. 

The corresponding equation of motion is given by 
\begin{gather}
\begin{split}
\partial_t \hat{\rho}_{1, \alpha  l s}^{\sigma} &\simeq  -i\mathcal{L} \hat{\rho}_{1, \alpha  l s}^{ \sigma}  -\omega_{\alpha \sigma l} \hat{\rho}_{1, \alpha  l s}^{\sigma}\\
&-i \Gamma \sum'_{\substack{s_1 s_2 s_{3} \\  \sigma_1 \sigma_2 \sigma_{3} }}  c^{\alpha}_{ j j_1  j_2 j_3}  \left[ \rho d^{ \bar{\sigma}_{3}}_{s_{3}}d^{\bar{\sigma}_{2}}_{s_{2}} d^{\bar{\sigma}_{1}}_{s_{1}} h_{\alpha l}^*  - d^{\bar{\sigma}_1}_{s_1} d^{\bar{\sigma}_2}_{s_2} d^{\bar{\sigma}_3}_{s_3} \rho h_{\alpha l} \right]  ,
\end{split} \\
\begin{split}
\partial_t \hat{\rho}_{2, \alpha  l s}^{\sigma} &\simeq  -i\mathcal{L} \hat{\rho}_{2, \alpha  l s}^{ \sigma}  -\omega_{\alpha \sigma l} \hat{\rho}_{2, \alpha  l s}^{\sigma} \\
&-i \Gamma \sum'_{\substack{s_1 s_2 s_{3} \\  \sigma_1 \sigma_2 \sigma_{3} }}  c^{\prime \alpha}_{ j_1 j j_2 j_3}  \left[d^{ \bar{\sigma}_2}_{s_{2}} d^{ \bar{\sigma}_3}_{s_{3}} \rho  d^{ \bar{\sigma}_{1}}_{s_{1}}   -  d^{ \bar{\sigma}_1}_{s_{1}}\rho  d^{ \bar{\sigma}_3}_{s_{3}} d^{ \bar{\sigma}_{2}}_{s_{2}} \right] \hat{h}_{\alpha l},
\end{split} 
\label{2p-order_eom}
\end{gather}
including the time-local approximation on $\hat{C}^{(2)}$ that we mentioned earlier and neglecting higher order terms. 
The summation is restricted to the combinations that conserve charges and spins; $\sigma + \sum_{i=1}^3\sigma_i = 0$, and $\sigma s + \sum_{i=1}^3 \sigma_i s_i= 0$. This is because $\hat{C}^{(2)}$ is only nonzero when these conditions are satisfied, as is known from Luttinger liquid theory. \cite{gogolin2004bosonization} The coefficient $c^{(\prime), \alpha}_{j j_1 j_2 j_3 }$ takes one of the values 
\begin{equation}
0, \ 2^{1-Y_{\alpha}}-1, \ 2^{Y_{\alpha}}-2, \ 2^{Y_{\alpha}-1}-1, \ 2^{-Y_{\alpha}}[1 - (-1)^{1-Y_{\alpha} }] ,
\label{C2_coeff}
\end{equation}
depending on the indices (see the explicit form in Appendix C). As expected, they vanish in the noninteracting limit $Y_{\alpha} = 1$. 
Please note that, in the following, we ignore the effect of higher orders of the $\hat{\rho}$ operators by truncating the hierarchy at this point, that is at the first tier of the $\hat{\rho}$ operators. 
This treatment is not, per se, systematic but allows us to estimate the leading order effect. As it turns out to be marginal in the parameter 
regime that we consider, we assume that higher order $\hat{\rho}$ contributions are even smaller and, therefore, negligible. In that sense, 
our results that we present in Sec.\ \ref{results} can be considered to be converged.

We are thus left with the infinite hierarchy of equations of the standard $\rho$ operators. We truncate it according to the arguments in Ref.~\onlinecite{Hartle2013}, where the importance of a $n (\geq 2)$th tier ADO, $\rho_\mathbf{j}$, is estimated by assigning it 
the following amplitude 
\begin{equation}
\prod_{k=1}^n  \left(   \frac{\left| h_{j_k} \right|}{\Re \omega_{j_k} \sum_{p=1}^{k-1} \Re \omega_{j_p}} \right).
\label{amplitude}
\end{equation}
In our calculations, we then consider only operators that have amplitudes larger than a given threshold value $A_\text{th}$. Convergence is achieved by reducing the threshold value systematically. In addition to the Grassmann and hermite natures of AIF's, this preselection greatly reduces the number of ADO's. Thus, we can obtain numerically converged, exact results for the ADO's of $\rho$ type (as was shown, for example, by a direct comparison to quantum Monte Carlo simulations only recently).\cite{Hartle2015} The $\hat{\rho}$ hierarchy is truncated at the first tier, ignoring higher order terms which correspond to $\mathcal{O}(\Gamma^{4})$. Since $n$-particle connected correlation functions scale as $\mathcal{O} (\Gamma^{n})$ and 
we typically obtain converged results already at second or third order for high enough temperatures, the truncation of the $\hat{\rho}$ hierarchy goes well along the lines of our convergence criterion. 
In the next section, we will show that the contributions from $\hat{C}^{(2)}$ have indeed only a very minor effect on the current-voltage characteristics. Therefore, we believe that essential correlation effects in the leads are included in our scheme. Yet, it is an important open problem how to go beyond the limitations and approximations of our scheme in order to get a systematic improvement of our results.

\subsection{Rate equations}
In the next section, we compare the HQME results with rate equation results. The latter contains only the lowest order of the hybridization expansion outlined above, ignoring, in addition, non-Markovian memory effects. We will see how these two assumptions affect the transport properties. The rate equations are derived for the weak dot-lead tunneling with Born-Markov-secular approximations:
\begin{equation}
\begin{split}
\dot{\rho} &=  - \int^{\infty}_0 d\tau  \sum_{\alpha s } \Big\{ \\
&\left[d_s(t) d^{\dagger}_s (t-\tau) \rho(t) - d^{\dagger}_s(t-\tau) \rho(t) d_s(t) \right] C^{+}_{\alpha }(\tau)  \\
&+ \left[d^{\dagger}_s(t) d_s (t-\tau) \rho(t) - d_s(t-\tau) \rho(t) d^{\dagger}_s(t) \right] C^{-}_{\alpha }(\tau) \\
&+ \text{h.c.} \Big\}.
\end{split}
\end{equation}
Using the diagonal form of the density matrix, 
\begin{equation}
\rho = P_0 \ket{0}\bra{0}+ \frac{P_{1}}{2}\left(  \ket{\uparrow}\bra{\uparrow} +  \ket{\downarrow}\bra{\downarrow} \right) + P_2\ket{\uparrow \downarrow}\bra{\uparrow \downarrow},
\label{population}
\end{equation}
we obtain the rate equations
\begin{equation}
\begin{split}
\dot{P}_0  &= \sum_{\alpha} \left[ -2 P_0 \mathcal{R}^{\alpha}_{01} +P_1 \mathcal{R}^{\alpha}_{10} \right], \\
\dot{P}_1  &= \sum_{\alpha} \left[ -P_1 \left( \mathcal{R}^{\alpha}_{10} + \mathcal{R}^{\alpha}_{12} \right)  + 2 P_0 \mathcal{R}^{\alpha}_{01} + 2 P_2 \mathcal{R}^{\alpha}_{21} \right], \\
\dot{P}_2  &= \sum_{\alpha} \left[ -2 P_2 \mathcal{R}^{\alpha}_{21} +P_1 \mathcal{R}^{\alpha}_{12} \right].
\end{split}
\end{equation}
The coefficients $\mathcal{R}^{\alpha}_{pq}$ are given by Laplace transforms of the correlation function $C^{\alpha}_0(\tau) \equiv C^{\sigma}_{\alpha } (\tau) e^{ -i \sigma \mu \tau}$ as 
\begin{equation}
\mathcal{R}^{\alpha}_{pq} = 2 \Re \int_{0}^{\infty} d \tau e^{- i \omega_{pq} \tau} C^{\alpha}_0(\tau) ,
\end{equation}
with $\omega_{01} = -\omega_{10} = \epsilon - \mu_{\alpha}$ and $\omega_{12} = -\omega_{21} = \epsilon + U - \mu_{\alpha}$. The current from lead $\alpha$ is 
\begin{equation}
\langle I_{\alpha} (t) \rangle = 2 P_0 \mathcal{R}_{01}^{\alpha} + P_1 \mathcal{R}_{12}^{\alpha} - P_1 \mathcal{R}^{\alpha}_{10} - 2 P_2 \mathcal{R}^{\alpha}_{21}.
\end{equation}
For the steady state, we find
\begin{multline}
\langle I_{\alpha} (t) \rangle = \frac{2e}{R} \Big[ \mathcal{R}_{21} \left(\mathcal{R}^{\alpha}_{01}\mathcal{R}^{\bar{\alpha}}_{10}-\mathcal{R}^{\bar{\alpha}}_{01} \mathcal{R}^{\alpha}_{10} \right) \\
 + \mathcal{R}_{01} \left(\mathcal{R}^{\alpha}_{12}\mathcal{R}^{\bar{\alpha}}_{21}-\mathcal{R}^{\bar{\alpha}}_{12} \mathcal{R}^{\alpha}_{21} \right) \Big],
\end{multline}
where $\mathcal{R}_{ij} = \sum_{\alpha} \mathcal{R}_{ij}^{\alpha}$, and 
\begin{equation}
R = \mathcal{R}_{10} \mathcal{R}_{21} + 2  \mathcal{R}_{01}  \mathcal{R}_{21} +  \mathcal{R}_{01}  \mathcal{R}_{12} .
\end{equation}


\section{Results}
\label{results}
\subsection{Weak coupling regimes}

\begin{table}[tb!]
\centering
\begin{ruledtabular}
\caption{Parameters used in our simulations. Note that these parameters are borrowed from molecular systems, but 
can be easily scaled to describe, for example, semiconductor heterostructeres, etc.}
\begin{tabular}{cccccc}
$W$ & $\Gamma$ & $U$ & $l_\text{max}$ & $A_\text{th}$  \\ \hline
4.0eV & 0.001eV &  0.25eV & $ 2500$ & $5.0\times10^{-10}$
\end{tabular}
\label{parameters}
\end{ruledtabular}
\end{table}

\begin{figure}[tb]
  \includegraphics[width=\columnwidth]{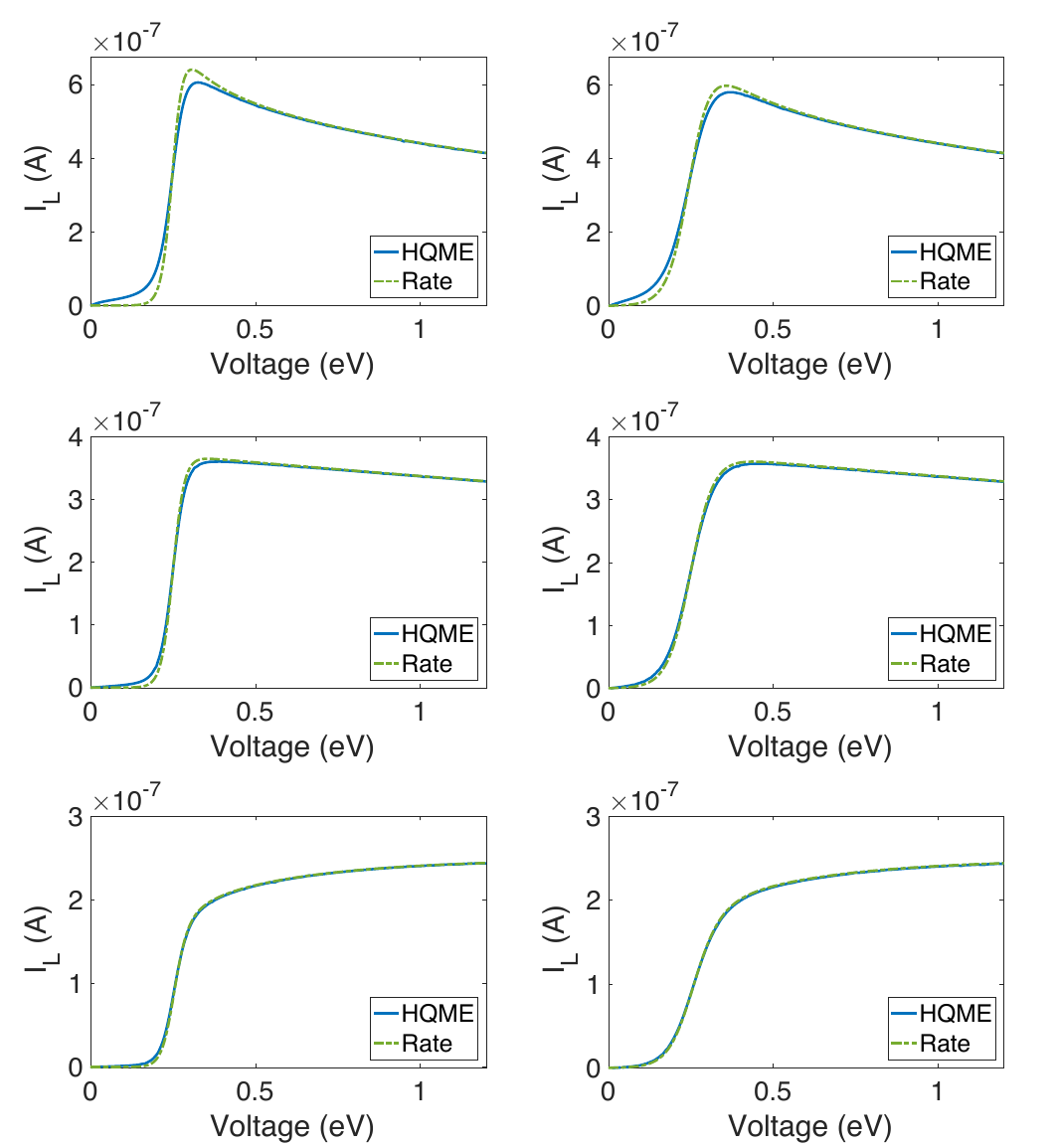}
\caption{Current-voltage characteristics for the quantum dot at the charge-symmetric point ($\epsilon=-U/2$), for 
various interaction parameters of the Luttinger liquid leads ($Y=0.8/1/1.2$ in the upper/middle/lower panels) and temperatures ($T=100/200$\,K in the left/right panels). Solid (dashed) lines are obtained by HQME (rate equations).}\label{IV_1}
\end{figure}
In this section, we show the current-voltage ($I-V$) characteristics of the model outlined in Sec.~\ref{model}. For simplicity, we consider symmetric leads; $W$, $\beta$, and $Y$ are independent of $\alpha$. The chemical potentials are taken in the standard form $\mu_{L} = -\mu_{R} = V/2$, corresponding to the choice of symmetric leads. We consider two scenarios where the quantum dot is at the charge symmetric point, $\epsilon = - U/2 $, and where it is unpopulated at zero bias, $\epsilon = U/2$. Other parameters of the model are given in Table \ref{parameters}.

In Fig.~\ref{IV_1}, we plot the current-voltage characteristics of a charge-symmetric quantum dot for three cases that are relevant, for example, for carbon nanotubes:\cite{Elste2011} attractive leads ($Y = 0.8$, upper panels), noninteracting leads ($Y=1$, middle panels), and repulsive leads ($Y= 1.2$, lower panels) for $T=100$\,K (left panels) and $T=200$\,K (right panels). In the attractive cases ($Y = 0.8 $), we observe a pronounced NDR above the Coulomb-blockade threshold at $V/2 = |\epsilon| = |\epsilon + U|$. This NDR is closely related to the peak in the single-particle tunneling density of states (see Fig.\ \ref{fig:psd}) and can also appear for noninteracting electrodes but attractive electron-electron interactions at the terminal site, \cite{Dubi2013} 
or slightly more general cases. \cite{Matveev1992} It appears quenched in the HQME results as compared to the rate equation results due to the effect of broadening or the hybridization with the leads. We also find that higher order or cotunneling processes which are not included in the rate equation results enhance the low bias conductance. The enhancement is strongest for the attractive case. This behavior can also be understood in terms of the peaked tunneling density of states (cf.\ Fig.\ \ref{fig:psd}), which is larger around the chemical potential 
in the attractive case as compared to the noninteracting or repulsive one. When $Y = 1$, the lead is noninteracting, and thus our formalism is numerically exact within the model. We find that due to the finite band width, $W = 4.0$ eV, the current has a small negative slope or NDR at high voltages $V \geq 1$. For repulsive leads ($Y = 1.2$), the curves monotonically increase in the range of voltages that we consider, and the deviation from the rate equations is negligible. This is because the density of states near the Fermi energy depends on the interaction parameter $Y$ with an explicit factor $W^{-Y}$ as seen in Eq.~\eqref{psi_corr}. Thus, effectively, the tunneling density of states is suppressed for repulsive interactions. This means that the perturbation parameter $\Gamma/T$ is also smaller and, thus, the differences between HQME and rate equations are also smaller.

\begin{figure}[!tb]
  \includegraphics[width=\columnwidth]{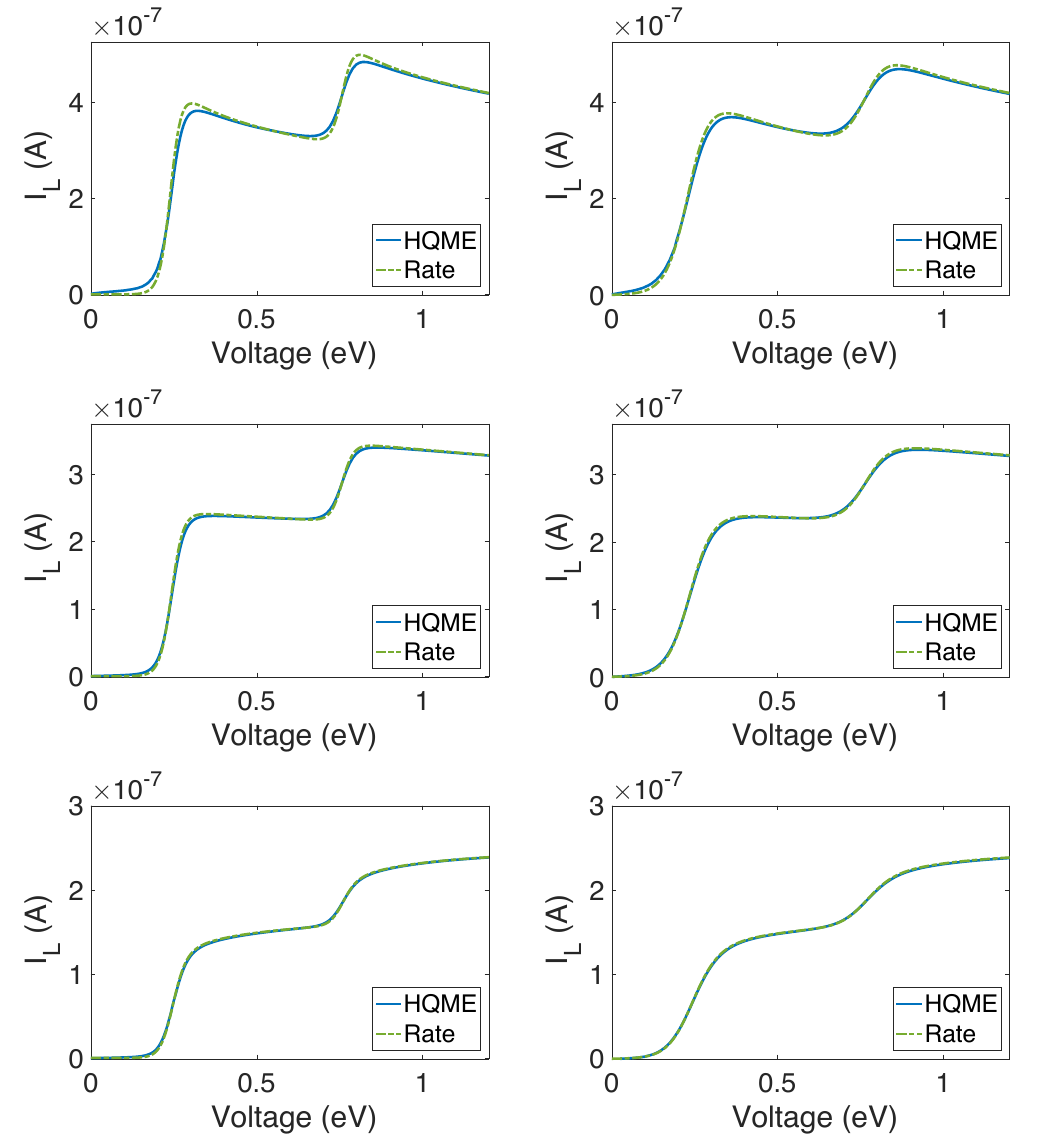}
\caption{Current-voltage characteristics for the quantum dot far away from the charge-symmetric point ($\epsilon=U/2$), for 
various interaction parameters of the Luttinger liquid leads ($Y=0.8/1/1.2$ in the upper/middle/lower panels) and temperatures ($T=100/200$\,K in the left/right panels). Solid (dashed) lines are obtained by HQME (rate equations).}\label{IV_2}
\end{figure}

Fig.~\ref{IV_2} shows the current-voltage characteristics for a quantum dot that is unpopulated at zero bias, $\epsilon = U/2$, i.e., far away from the charge symmetric point. A similar situation has been considered, for example, in Ref.~\onlinecite{Elste2011}. Here, in contrast to the charge-symmetric case, we observe two steps at $V/2=|\epsilon|$ and $V/2=|\epsilon+U|$. In the charge-symmetric case, they appear at the same voltages. Both steps are followed by a decrease of the current level in the attractive case $Y=0.8$, similar as in the charge-symmetric case. The effect is again more pronounced in the rate equation solutions due to broadening that is missing in the rate equation results. 
Similar to the charge-symmetric case, the effect of cotunneling processes at low bias voltages is more pronounced at lower temperatures (see the upper left and upper right panels). It is generally less pronounced in the charge-non-symmetric case, because the single-particle levels at $\epsilon+U$ are farther away from the chemical potentials, that is the associated probability for virtual excitations is lower as compared to the charge-symmetric case.

\begin{figure}[tb]
  \includegraphics[width=0.8 \columnwidth]{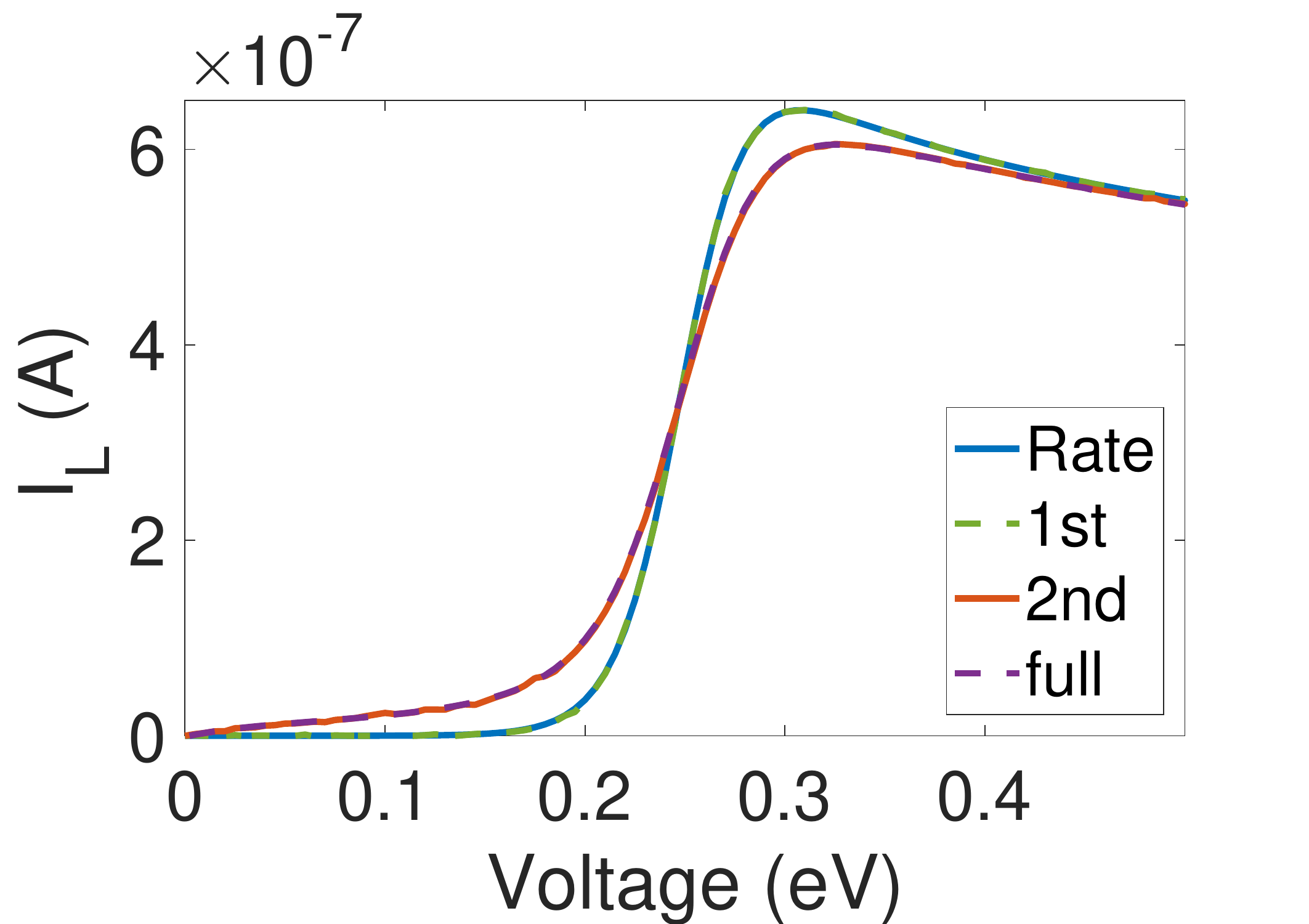}
\caption{Comparison of current-voltage characteristics for the quantum dot at the charge-symmetric point ($\epsilon=-U/2$) for 
$Y=0.8$ and $T=100$\,K obtained with rate equations (blue line), full first order (dashed green line), second order (orange line) and the converged HQME results (dashed purple line).}\label{IV_comp}
\end{figure}

Differences between the rate equation and HQME results reveal the effect of higher order processes, including second and beyond-second order processes. Some more insights can be obtained by comparing other, approximate solutions. The role of the Markov approximation, for example, can be assessed by comparing rate equation results with a first-tier truncation of the HQME. Such a comparison is depicted by the blue and green lines in Fig.\ \ref{IV_comp}. The negligible differences between the two curves shows that the Markov approximation is well justified in the parameter regime that we consider. Moreover, comparing a second-tier truncation of the HQME with the full converged result (yet, including the approximations with respect to multi-particle correlations; see Sec.\ \ref{method}) allows us to reveal the effect of beyond-second order processes (cf.\ orange and purple lines in Fig.\ \ref{IV_comp}). They also turn out to be negligible. 
Please note that a different behavior can be expected at low temperatures, especially when, for example, Kondo correlations emerge \cite{Hartle2015}.

\subsection{Strong coupling regimes}
So far, we looked at the cases where interactions of the leads are relatively weak ($|Y-1| \lesssim 20\%$), since our decomposition scheme, 
Eq.~\eqref{C_approx}, converges well only for these cases. However, in this regime, the effect of the two-particle correlation functions $\hat{C}^{(2)}$ is quite small since the order of the corresponding ADO's $\hat{\rho}$ is
\begin{equation}
\mathcal{O} (\hat{\rho} )\sim \left( \frac{\Gamma}{T} \right)^2  \left(\frac{T}{W} \right)^{Y} c_{j_1j_2j_3j_4}.
\end{equation}
This estimate is motivated by Eq.~\eqref{2p-order_eom}; in the steady state, the amplitude of the $\hat{\rho}_{1,\alpha s}^{\sigma}$ and $\hat{\rho}_{2,\alpha s}^{\sigma}$ operators 
scale with $\Gamma h$ and $\Gamma \hat{h}$ respectively and with $1/\omega_l$, including an additional factor that vanishes as the interaction strength decreases. Thus, in the above weak coupling regimes, we find the effect of $\hat{\rho}$ ignorable. The complete treatment of both $\hat{\rho}$ and regular ADO's at higher tiers is desirable but not possible within our current scheme. That being said, we would like to illustrate the effects of two-particle correlation functions for relatively strong couplings realized by very high temperatures ($Y= 0.5$, $\Gamma = 0.02$ eV, and $T=4000$\,K, other parameters are the same as in the previous subsection), and stop the hierarchy of the regular ADO's at the first tier. This allows us to disentangle the higher order tunneling effect from regular ADO's and from two-particle correlations that enter via the $\hat{\rho}$ operators.

\begin{figure}[tb]
  \includegraphics[width=0.8\columnwidth]{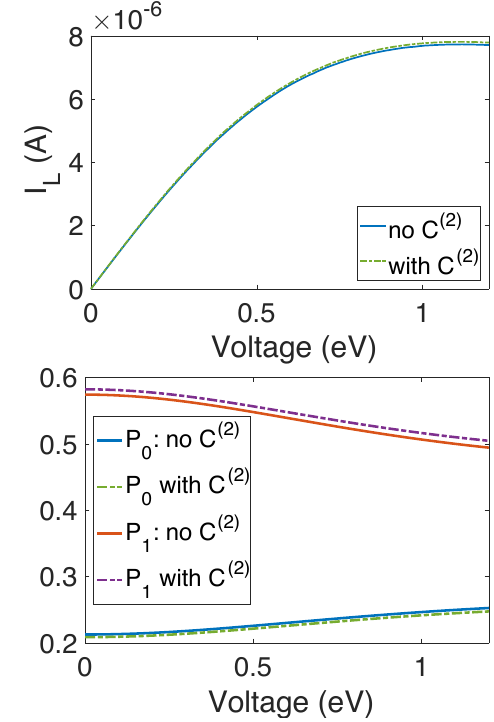}
\caption{The top panel shows the current-voltage characteristics with and without contributions of two-particle correlations $C^{(2)}$, i.e., $\hat{\rho}$ ADO's. The bottom panel shows the dot populations, Eq.~\eqref{population}, with and without contributions of two-particle correlations $C^{(2)}$. }
\label{rho_hat}
\end{figure}

Fig.~\ref{rho_hat} shows the current-voltage characteristics and dot populations with and without the $\hat{\rho}$ contributions. For this extreme case, we find that the effect of two particle correlation functions is still negligible for the current level ($\sim 2 \%$), while the dot populations are slightly more affected ($\sim 5\%$). The current is slightly enhanced by two-particle correlations. The probability of having a single particle on the dot is also increased, while the ones of having no particle or two particles are reduced. 

\section{Conclusions}
\label{conclusion}

We presented a generalized hierarchical quantum master equation technique to describe transport through an interacting quantum dot or molecule that is coupled to interacting, semi-infinite, Luttinger liquid leads. Our method paves the way towards a numerically exact description of this transport problem. This, however, is not fully achieved yet. While the results can be converged with respect to single-particle correlations in the leads [where we included contributions up to fifth order, $\mathcal{O}(\Gamma^5)$, in order to obtain fully converged results], multi-particle correlations can be included only in leading order, i.e., $\mathcal{O}(\Gamma^2)$. A systematic improvement of the latter approximation represents the next step. Nevertheless, we were able to present well based arguments and numerical results, which show that these contributions are negligible for the parameters that we considered, in particular the high temperature regime. Thus, assuming that a hybridization expansion converges, our method can give results that should be very close to the exact result. 

As a test case, we considered transport through a quantum dot or molecule that can be described by a single-level Anderson impurity model. We corroborated previous results on a mechanism for negative differential resistance that can be associated with the power-law singularity of an attractive Luttinger liquid. In the repulsive case, the power-law singularity has the opposite effect and even overrides the NDR effect originating from a finite band width in the leads. Higher order effects smear the NDR effects slightly due to broadening. In the low bias regime, cotunneling effects enhance the conductivity significantly. While this is well known, we were able to show that beyond-second order processes do not change these effects significantly. This justifies, inter alia, a treatment by second-order perturbation theory. Our statements and findings are restricted to the high temperature regime and the reduced complexity of our model. More complex quantum dot structures, including, for example, also spin-orbit interactions or the coupling to vibrational degrees of freedom, may show a richer behavior. \\

\section{Acknowledgements}
We thank K.\ Sch\"onhammer for helpful discussions and comments. 
RH is supported by Deutsche Forschungsgemeinschaft (DFG) under grant No.\ HA 7380/1-1. 
JO and LM are supported by the Deutsche Forschungsgemeinschaft through the SFB 925 and the Hamburg Centre for Ultrafast Imaging, and from the Landesexzellenzinitiative Hamburg, supported by JoachimHerzStiftung.

\onecolumngrid
\appendix
\section{General formula of the influence functional $\mathcal{F}$}
\label{appA}

For the simple coupling Hamiltonian $H_\text{tun}^{\alpha} (t)$, the influence functional is a simple sum $\mathcal{F} = \sum_{\alpha = L, R} \mathcal{F}_{\alpha}$. Thus, in the following, we focus on one of the terms $\mathcal{F}_{\alpha}$ and omit the index $\alpha$. The situations may be different for excitonic coupling, where the lead operator is transformed into an operator that has bosonic fields from both leads,
\begin{equation}
\tilde{f}_{\alpha s} = f_{\alpha s} e^{-\sum_{\alpha' } \frac{1}{u_c} \sqrt{\frac{K_c}{2}} V_{\alpha'} \phi_{\alpha' c R}}.
\end{equation}
However, due to charge-neutrality, the decoupling of $\mathcal{F}_L$ and $\mathcal{F}_R$ still holds. 

To calculate the influence exponent $\Phi=-\log \mathcal{F}$ (see Appendix\ \ref{appB}), we expand the exponentials in Eq.~\eqref{FV} and take the average over the leads degrees of freedom. The resulting terms include both disconnected and connected graphs in the replica sense, and we will subtract the disconnected graphs later. A $2N$th order contribution is (odd orders always vanish)
\begin{multline}
\mathcal{F}^{(2N)}= (-1)^N \sum_{k= 0}^{2N} (-1)^k    \sum_{\substack{s \cdots s_k \\ s'_1 \cdots s'_{2N-k} } } \sum_{\substack{ \sigma_1 \cdots \sigma_k \\ \sigma'_1 \cdots \sigma'_{2N-k} }} \int_{t_0}^t d\tau_1 \cdots \int_{t_0}^{\tau_{k-1}} d\tau_k   \int_{t_0}^t d\tau'_1 \cdots \int_{t_0}^{\tau'_{2N-k-1}} d\tau'_{2N-k} \\
\times \tilde{\xi}^{\bar{\sigma}_1}_{s_1}(\tau_1) \cdots  \tilde{\xi}^{\bar{\sigma}_k}_{s_k}(\tau_k) \tilde{\xi}^{\prime, \bar{\sigma}'_{2N-k}}_{s'_{2N-k}}(\tau'_{2N-k}) \cdots  \tilde{\xi}^{\prime, \bar{\sigma}'_1 }_{s'_1}(\tau'_1)  \langle f^{\sigma'_{2N-k}}_{s'_{2N-k}} (\tau'_{2N-k})   \cdots   f^{\sigma'_1}_{s'_1} (\tau'_1)  f^{\sigma_1}_{s_1} (\tau_1) \cdots  f^{\sigma_k}_{s_k} (\tau_k)   \rangle_{L+R} ,
\label{F2N}
\end{multline}
where we have already used the fact that the average over the leads variables requires the charge and spin conservations as follows. The lead correlation functions can be calculated via bosonization, noting $f_{s}^{\sigma} = a \bar{\mathcal{T}} \bar{\psi}_s^{\sigma}$,
\begin{multline}
\langle f^{\sigma_{1}}_{s_1} (\tau_1)   \cdots   f^{\sigma_{2N}}_{s_{2N}} (\tau_{2N}) \rangle_{L+R} =
 \Gamma^N \delta \left( \sum_i \sigma_i \right) \delta \left( \sum_i \sigma_i s_i \right)  \left( \prod_{i}^{2N} \eta_{s_i} \right)   \\
\times e^{i  \mu \sum_{i} \sigma_i \tau_i } e^{- \sum_{i<j} \sigma_i \sigma_j D_c(\tau_i - \tau_j)} e^{- \sum_{i<j} \sigma_i s_i \sigma_j s_j D_s(\tau_i - \tau_j)},
 \label{f_corr}
\end{multline}
where $\eta$ is the Klein factor, and $D_{c/s}$ are asymptotically
\begin{equation}
D_{c/s}(t) \sim -\frac{1}{2K_{c/s}} \ln \left\{ \frac{i \beta v_F}{a \pi} \sinh \left[ \frac{\pi (t-i \delta)}{\beta} \right] \right\}
\label{D_cs}
\end{equation}
with the Luttinger parameters $K_{c/s}$. We assume $K_s = 1$, and $K_c$ can be either an originally repulsive value ($K_c < 1$) or an effective attractive value through excitonic coupling as in Eq.~\eqref{Y_renorm}.

Equations.~\eqref{F2N}, \eqref{f_corr}, and \eqref{D_cs} give a formally exact expression of the influence functional for interacting Luttinger liquid leads.

\section{Influence exponent $\Phi$}
\label{appB}

To derive the HQME, we need to calculate the influence exponent $\Phi \equiv - \log \mathcal{F}$. The cumulant expansion of $\Phi$ can be calculated by using the replica trick
\begin{equation}
\log \mathcal{F} = \lim_{n \rightarrow 0} \frac{d \mathcal{F}^n}{dn}.
\end{equation}
From the series expansion of $\mathcal{F}$ in Eq.~\eqref{FV_exp}, we find that $\mathcal{F}^{n}$ scales as
\begin{equation}
\mathcal{F}^n \sim 1 + n \int C^{(1)} \tilde{\xi}^2 + n\int C^{(2)}\tilde{\xi}^4 + \frac{n(n-1)}{2} \left[ \int C^{(1)} \tilde{\xi}^2 \right]^2  \dots,
\end{equation}
In the limit $n \rightarrow 0$, only the terms proportional to $n$ survive in $\log \mathcal{F}$. We will show an explicit formula up to $C^{(2)} \sim \mathcal{O}(\Gamma^2)$. Higher-order contributions can be derived in the same manner. 

\subsection{2nd order}
There are six terms from $\mathcal{F}^{(2)}$,
\begin{equation} \begin{split}
\Phi^0 &\equiv \sum_{s, s'} \int_{t_0}^{t} d\tau \int^{\tau}_{t_0} d \tau' \left[ \tilde{\xi}_{s} (\tau) \tilde{\xi}^{*}_{s'} (\tau') C^{+} (\tau -\tau') + \tilde{\xi}_{s}^{*} (\tau) \tilde{\xi}_{s'} (\tau') C^{-} (\tau -\tau') \right] \\
&+ \sum_{s s'} \int_{t_0}^{t} d\tau \int^{\tau}_{t_0} d \tau' \left[  \tilde{\xi}'_{s'} (\tau')  \tilde{\xi}_{s}^{\prime *} (\tau) C^{+*}(\tau -\tau') + \tilde{\xi}_{s'}^{\prime *}  (\tau' ) \tilde{\xi}_{s}' (\tau) C^{-*} (\tau -\tau') \right] \\
&- \sum_{s s'} \int_{t_0}^{t} d\tau \int^{t}_{t_0} d \tau' \left[  \tilde{\xi}_{s} (\tau) \tilde{\xi}_{s'}^{\prime *} (\tau') C^{-*} (\tau -\tau') +\tilde{\xi}_{s}^*  (\tau) \tilde{\xi}_{s'}' (\tau') C^{+*} (\tau -\tau') \right].
\end{split}\end{equation} 
Without interactions in the leads, this is the exact expression of the influence exponent \cite{Jin2008}, and all higher order contributions vanish.

\subsection{4th order}
The total fourth order contributions are
\begin{multline}
\Phi' \equiv  -  \sum_{k= 0}^{4}  (-1)^k \sum'_{j_1, \dots, j_k, j'_{4-k}, \dots, j'_1} \int_{t_0}^t d\tau_1 \cdots \int_{t_0}^{\tau_{k-1}} d\tau_k   \int_{t_0}^t d\tau'_1 \cdots \int_{t_0}^{\tau'_{3-k}} d\tau'_{4-k}  \\
  \times  \tilde{\xi}_{\bar{j}_1}(\tau_1) \cdots  \tilde{\xi}_{\bar{j}_k}(\tau_k) \tilde{\xi}'_{\bar{j}'_{4-k}}(\tau'_{4-k}) \cdots  \tilde{\xi}'_{\bar{j}'_1 }(\tau'_1)  \hat{C}^{(2)}_{j'_{4-k}\cdots j_k}(\tau'_{4-k}, \cdots, \tau_k) 
\end{multline}
where $j = ( \sigma, s)$, $\bar{j} = (\bar{\sigma}, s)$, and $\sum'$ indicates the summation with the charge and spin conservations. The connected correlation functions are defined as 
\begin{multline}
\hat{C}^{(2)}_{j_1 \cdots j_4}(\tau_1, \cdots, \tau_4)
= \langle f^{\sigma_1}_{s_1} (\tau_1)   f^{\sigma_2}_{s_2} (\tau_2)  f^{\sigma_3}_{s_3} (\tau_3)    f^{\sigma_4}_{s_4} (\tau_4)   \rangle_{L+R} 
- \langle f^{\sigma_1}_{s_1} (\tau_1)   f^{\sigma_2}_{s_2} (\tau_2)\rangle_{L+R} \langle  f^{\sigma_3}_{s_3} (\tau_3)    f^{\sigma_4}_{s_4} (\tau_4)   \rangle_{L+R}  \\
+ \langle f^{\sigma_1}_{s_1} (\tau_1)    f^{\sigma_3}_{s_3} (\tau_3) \rangle_{L+R} \langle   f^{\sigma_2}_{s_2} (\tau_2)  f^{\sigma_4}_{s_4} (\tau_4)   \rangle_{L+R} 
- \langle f^{\sigma_1}_{s_1} (\tau_1) f^{\sigma_4}_{s_4} (\tau_4)   \rangle_{L+R} \langle   f^{\sigma_2}_{s_2} (\tau_2)  f^{\sigma_3}_{s_3} (\tau_3)    \rangle_{L+R}
\end{multline}
They obey the symmetry relations 
\begin{equation}
[\hat{C}^{(2)}_{j_1j_2j_3j_4} (\tau_1, \tau_2, \tau_3, \tau_4)]^*  = \hat{C}^{(2)}_{\bar{j}_4 \bar{j}_3 \bar{j}_2 \bar{j}_1} (\tau_4, \tau_3, \tau_2, \tau_1) .
\label{C2_symm}
\end{equation}
Using the following identity for $s_{ij} \equiv \sinh(t_{i} - t_j)$,
\begin{equation}
s_{12} s_{34} + s_{23} s_{14} = s_{13} s_{24},
\end{equation}
one can prove that $\hat{C}^{(2)}$ indeed vanishes in the noninteracting limit, $Y = 1$, for $\delta \rightarrow 0$.

The influence exponent $\Phi'$ follows
\begin{equation}
\partial_t \Phi' = -i \mathcal{A}_{\bar{j}} \left(\hat{\mathcal{B}}_{1,j}+  \hat{\mathcal{B}}_{2,j}\right),
\end{equation}
where
\begin{equation}
\begin{split}
\hat{\mathcal{B}}_{1, j}  &=  -i\int_{t_0}^t d\tau_1  \int_{t_0}^{\tau_{1}} d\tau_{2} \int_{t_0}^{\tau_{2}} d\tau_{3} \sum'_{j_1 j_2 j_{3} }  \tilde{\xi}'_{\bar{j}_{3}}(\tau_{3})\tilde{\xi}'_{\bar{j}_{2}}(\tau_{2}) \tilde{\xi}'_{\bar{j}_{1}}(\tau_{1})  \hat{C}^{(2)}_{j_3 j_2 j_1 j} (\tau_3, \tau_2, \tau_1, t) \\
&+ i  \int_{t_0}^t d\tau_1 \int_{t_0}^{\tau_{1}} d\tau_2 \int_{t_0}^{\tau_{2}} d\tau_3    \sum'_{j_1 j_2 j_{3}}    \tilde{\xi}_{\bar{j}_1}(\tau_1) \tilde{\xi}_{\bar{j}_2}(\tau_2) \tilde{\xi}_{\bar{j}_3}(\tau_3)    \hat{C}^{(2)}_{j j_1  j_2 j_3} (t, \tau_1, \tau_2, \tau_3), \\
\hat{\mathcal{B}}_{2, j} & = -i\int_{t_0}^t d\tau_1 \int_{t_0}^{\tau_{1}} d\tau_{2} \int_{t_0}^{t} d\tau'_{1} \sum'_{j_1 j_2  j'_{1} } \tilde{\xi}_{\bar{j}_1}(\tau_1) \tilde{\xi}_{\bar{j}_2}(\tau_2) \tilde{\xi}'_{\bar{j}'_{1}}(\tau'_{1})   \hat{C}^{(2)}_{ j'_1  j j_1 j_2} (\tau'_1,  t, \tau_1, \tau_2)\\
&+i \int_{t_0}^t d\tau_1 \int_{t_0}^{\tau_{1}} d\tau_{2} \int_{t_0}^{t} d\tau'_{1} \sum'_{j_1 j_2  j'_{1} }   \tilde{\xi}_{\bar{j}'_1}(\tau'_1)   \tilde{\xi}'_{\bar{j}_{2}}(\tau_{2})  \tilde{\xi}'_{\bar{j}_{1}}(\tau_{1})   \hat{C}^{(2)}_{ j_2 j_1 j j'_1} ( \tau_2, \tau_1, t, \tau'_1).
\end{split}
\end{equation}
$\mathcal{A}_{\bar{j}}$ is a superoperator defined in Eq.~\eqref{ABC}. Using the symmetry of the correlation functions, Eq.~\eqref{C2_symm}, we can prove that $\hat{\mathcal{B}}_{1(2),s}^{\bar{\sigma}}  = [\hat{\mathcal{B}}_{1(2),s}^{\sigma}]^{\dagger}$, thus the corresponding ADO is Hermitian.

\section{Approximation of $\hat{C}^{(2)}$}
\label{appC}

\label{approx_C4}
\begin{table}[!tb]
\begin{center}
\caption{Coefficients for Eqs.~\eqref{C4_approx_1} and \eqref{C4_approx_2}. $j=1, 2, 3, 4$ corresponds to $\{\sigma s \} = \{+\uparrow, +\downarrow, -\uparrow, -\downarrow\}$. $c_1 = 2^{1-Y}-1$, $c_2 = 2^{Y}-2$, $c_3 = 2^{Y-1}-1$, and $c_4 = 2^{-Y}[1 - (-1)^{1-Y }]$.}
\label{cijkl}
\begin{tabular}{cccccc}
$j_1$ & $j_2$ & $j_3$ & $j_4$ & $c_{j_1j_2j_3j_4}$ & $c'_{j_1j_2j_3j_4}$ \\
\hline
 1 & 1 & 3 & 3 & $c_1$ & $c_4$\\
 1 & 2 & 3 & 4 & 0  & $c_4$ \\
 1 & 2 & 4 & 3 & $c_1$ & 0\\
 1 & 3 & 1 & 3 & $c_2$ & 0\\
 1 & 3 & 2 & 4 & $c_3$ & 0\\
 1 & 3 & 3 & 1 & $c_1$ & $c_4^*$\\
 1 & 3 & 4 & 2 & $c_1$ & 0\\
 1 & 4 & 2 & 3 & $c_3$ & 0\\
 1 & 4 & 3 & 2 & 0 & $c_4^*$\\
 2 & 1 & 3 & 4 & $c_1$ & 0\\
 2 & 1 & 4 & 3 & 0 & $c_4$\\
 2 & 2 & 4 & 4 & $c_1$ & $c_4$\\
 2 & 3 & 1 & 4 & $c_3$ & 0\\
 2 & 3 & 4 & 1 & 0 & $c_4^*$\\
 2 & 4 & 1 & 3 & $c_3$ & 0\\
 2 & 4 & 2 & 4 & $c_2$ & 0\\
 2 & 4 & 3 & 1 & $c_1$ & 0\\
 2 & 4 & 4 & 2 & $c_1$ & $c_4^*$\\
 3 & 1 & 1 & 3 & $c_1$ & $c_4^*$\\
 3 & 1 & 2 & 4 & $c_1$ & 0\\
 3 & 1 & 3 & 1 & $c_2$ & 0\\
 3 & 1 & 4 & 2 & $c_3$ & 0\\
 3 & 2 & 1 & 4 & 0 & $c_4^*$\\
 3 & 2 & 4 & 1 & $c_3$ & 0\\
 3 & 3 & 1 & 1 & $c_1$ & $c_4$\\
 3 & 4 & 1 & 2 & 0 & $c_4$\\
 3 & 4 & 2 & 1 & $c_1$ & 0\\
 4 & 1 & 2 & 3 & 0 & $c_4^*$\\
 4 & 1 & 3 & 2 & $c_3$ & 0\\
 4 & 2 & 1 & 3 & $c_1$ & 0\\
 4 & 2 & 2 & 4 & $c_1$ & $c_4^*$\\
 4 & 2 & 3 & 1 & $c_3$ & 0\\
 4 & 2 & 4 & 2 & $c_2$ & 0\\
 4 & 3 & 1 & 2 & $c_1$ & 0\\
 4 & 3 & 2 & 1 & 0 & $c_4$\\
 4 & 4 & 2 & 2 & $c_1$ & $c_4$\\
\end{tabular}
\end{center}
\end{table}

\begin{figure}[!tb]
  \includegraphics[width=0.9\columnwidth]{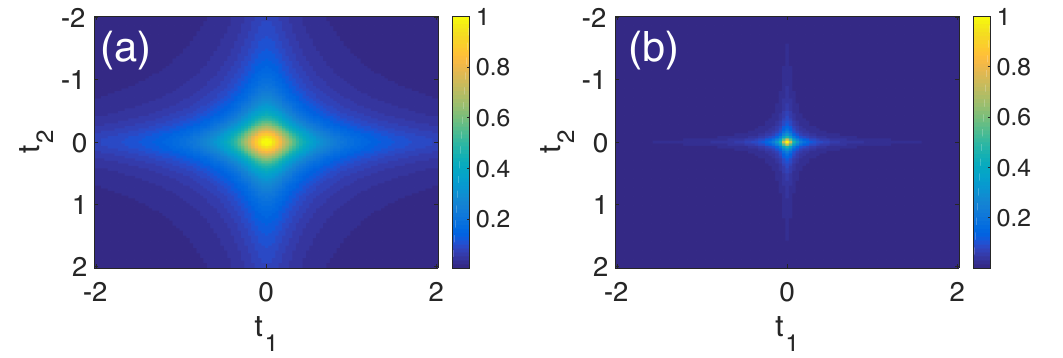}
\caption{$| \hat{C}^{(2)}(0,0,t_1,t_2) |$ at $T=200$ K for $Y=1.2$. (a) $W = 4$ eV and (b) $W=20$ eV. The correlation function decays exponentially. }\label{C4_LTA}
\end{figure}

To derive a closed set of HQME, we approximate the correlation functions $\hat{C}^{(2)}$ noting that this function is locally peaked in time when $\tau_1 = \tau_2 = \tau_3 = \tau_4$.\cite{Jang2002} To illustrate this, we plot the absolute value of a two-particle correlation function $\hat{C}^{(2)}(0,0,t_1,t_2)$ for $W=4$ eV and $20$ eV in Fig.~\ref{C4_LTA}. We see that this function rapidly decreases away from the origin. The decrease is even more pronounced as the bandwidth $W$ becomes larger. This motivates to approximate the function by just two time variables, e.g., $\hat{C}^{(2)} (\tau_1, \tau_2, \tau_3, \tau_4) \simeq \hat{C}^{(2)} (\tau_1, \tau_1, \tau_1, \tau_4)$. More precisely, we approximate the connected two particle correlation functions in $\hat{\mathcal{B}}$ as
\begin{equation}
\begin{split}
\hat{\mathcal{B}}_{1, j} &\simeq  -i\int_{t_0}^t d\tau \sum'_{j_1 j_2 j_{3}} \left[  \tilde{\xi}'_{\bar{j}_{3}}(\tau)\tilde{\xi}'_{\bar{j}_{2}}(\tau) \tilde{\xi}'_{\bar{j}_{1}}(\tau)  \hat{C}^{(2)}_{j_3 j_2 j_1 j} (\tau, \tau, \tau, t) - \tilde{\xi}_{\bar{j}_1}(\tau) \tilde{\xi}_{\bar{j}_2}(\tau) \tilde{\xi}_{\bar{j}_3}(\tau)    \hat{C}^{(2)}_{j j_1  j_2 j_3} (t, \tau, \tau, \tau) \right],\\
\hat{\mathcal{B}}_{2, j} &\simeq -i\int_{t_0}^t d\tau  \sum'_{j_1 j_2  j'_{1} } \left[ \tilde{\xi}_{\bar{j}_1}(\tau) \tilde{\xi}_{\bar{j}_2}(\tau) \tilde{\xi}'_{\bar{j}'_{1}}(\tau)   \hat{C}^{(2)}_{ j'_1  j j_1 j_2} (\tau,  t, \tau, \tau) -    \tilde{\xi}_{\bar{j}'_1}(\tau)   \tilde{\xi}'_{\bar{j}_{2}}(\tau)  \tilde{\xi}'_{\bar{j}_{1}}(\tau)   \hat{C}^{(2)}_{ j_2 j_1 j j'_1} ( \tau, \tau, t, \tau) \right].
\end{split}
\end{equation}
In the next step, we need to find a series expansion of the approximated two-particle correlation function by exponentials, whose specific form depends on the superindices $j$. Using spin and charge conservation, we can reduce $\hat{C}^{(2)}(t, \tau,\tau,\tau)$ into $C^{\sigma}$ in Eq.~\eqref{ABC} as 
\begin{equation}
\hat{C}^{(2)}_{j_1 j_2 j_3 j_4}(t, \tau, \tau, \tau) = c_{j_1j_2j_3j_4}   \Gamma  C^{\sigma_1} (t- \tau).
\label{C4_approx_1}
\end{equation}
The factors $c_{j_1 j_2 j_3 j_4}$ are listed in Table~\ref{cijkl}. $\hat{C}^{(2)}(\tau,\tau,\tau, t)$ is obtained from the symmetry, Eq.~\eqref{C2_symm}, and $c_{j_1 j_2 j_3 j_4} =  c_{\bar{j_4}\bar{j_3}\bar{j_2}\bar{j_1}} = c_{j_4 j_3 j_2 j_1}$. This leads to 
\begin{equation}
\hat{\mathcal{B}}_{1,j} =   -i  \Gamma  \int_{t_0}^t d\tau \sum'_{j_1 j_2 j_{3}} c_{ j j_1  j_2 j_3} \sum_l  \Big[  \tilde{\xi}'_{\bar{j}_{3}}(\tau)\tilde{\xi}'_{\bar{j}_{2}}(\tau) \tilde{\xi}'_{\bar{j}_{1}}(\tau) h_{l}^* - \tilde{\xi}_{\bar{j}_1}(\tau) \tilde{\xi}_{\bar{j}_2}(\tau) \tilde{\xi}_{\bar{j}_3}(\tau) h_l \Big] e^{-\omega_l^{\sigma}(t-\tau)} \equiv \sum_l \hat{\mathcal{B}}_{1, j, l}.
\end{equation}
The equation of motion for each frequency component is found to be
\begin{equation}
\partial_t \hat{\mathcal{B}}_{1,j, l } =  - \omega_l^{\sigma}\hat{\mathcal{B}}_{1,j, l } -i  \Gamma \sum'_{j_1 j_2 j_{3}}c_{ j j_1  j_2 j_3}   \left[  \xi'_{\bar{j}_{3}} \xi'_{\bar{j}_{2}} \xi'_{\bar{j}_{1}} h_{l}^*  - \xi_{\bar{j}_1} \xi_{\bar{j}_2} \xi_{\bar{j}_3} h_l \right].
\end{equation}
Similarly, we find in the wide band limit
\begin{equation}
\hat{C}^{(2)}_{j_1 j_2 j_3 j_4}(\tau,t, \tau, \tau) \simeq \Gamma^2 e^{i \mu \sigma_2 (t-\tau) } c'_{j_1 j_2 j_3 j_4}  \left[ \frac{\beta W}{\pi} \cosh \left(\frac{\pi  (t-\tau)}{\beta}\right) \sin \left(\frac{\pi }{\beta W}\right)\right]^{-Y} \equiv \Gamma  c'_{j_1 j_2 j_3 j_4}  C^{\sigma_2} (t-\tau),
\label{C4_approx_2}
\end{equation}
which can be decomposed into an exponential series as
\begin{equation}\begin{split}
 C^{\sigma} (t) &= \sum_{l = 0}^{\infty} \hat{h}_{l } e^{- \omega_{l }^{\sigma} t},\\
\hat{h}_{ l}&=  |\Gamma_{\alpha}| 2^{Y} \frac{(Y)_l}{l!} \left[ \frac{W  \beta}{\pi}  \sin \left(\frac{\pi }{\beta W}\right)  \right]^{-Y}.
\end{split}
\end{equation}
$ \omega_{l }^{\sigma}$ is the same as in Eq.~\eqref{C_approx}. Thus, we find
\begin{equation}
\hat{\mathcal{B}}_{2,j} = -i \Gamma \int_{t_0}^t d\tau  \sum'_{j_1 j_2  j'_{1} }      c'_{ j'_1  j j_1 j_2}\left[ \tilde{\xi}_{\bar{j}_1}(\tau)  \tilde{\xi}_{\bar{j}_2}(\tau)   \tilde{\xi}'_{\bar{j}'_1}(\tau) -  \tilde{\xi}_{\bar{j}'_1}(\tau)  \tilde{\xi}'_{\bar{j}_2}(\tau)  \tilde{\xi}'_{\bar{j}_1}(\tau) \right] \hat{h}_l e^{-\omega_l^{\sigma} (t-\tau)}\equiv \sum_l \hat{\mathcal{B}}_{2,j, l}.
\end{equation}
The equation of motion for each frequency component is
\begin{equation}
\partial_t \hat{\mathcal{B}}_{2,j, l } =  - \omega_l^{\sigma}\hat{\mathcal{B}}_{2,j, l } -i  \Gamma \sum'_{j_1 j_2 j'_{1}}c'_{  j'_1  j j_1 j_2}    \left[  \xi_{\bar{j}_1} \xi_{\bar{j}_2}  \xi'_{\bar{j}'_1}-  \xi_{\bar{j}'_1}\xi'_{\bar{j}_2}  \xi'_{\bar{j}_1}\right] \hat{h}_l.
\end{equation}
The resulting equations of motion for the ADO's are given by Eqs.~\eqref{2p-order_eom}. 

\twocolumngrid


\begin{thebibliography}{93}
\expandafter\ifx\csname natexlab\endcsname\relax\def\natexlab#1{#1}\fi
\expandafter\ifx\csname bibnamefont\endcsname\relax
  \def\bibnamefont#1{#1}\fi
\expandafter\ifx\csname bibfnamefont\endcsname\relax
  \def\bibfnamefont#1{#1}\fi
\expandafter\ifx\csname citenamefont\endcsname\relax
  \def\citenamefont#1{#1}\fi
\expandafter\ifx\csname url\endcsname\relax
  \def\url#1{\texttt{#1}}\fi
\expandafter\ifx\csname urlprefix\endcsname\relax\def\urlprefix{URL }\fi
\providecommand{\bibinfo}[2]{#2}
\providecommand{\eprint}[2][]{\url{#2}}

\bibitem[{\citenamefont{Kastner}(2000)}]{Kastner2000}
\bibinfo{author}{\bibfnamefont{M.~A.} \bibnamefont{Kastner}},
  \bibinfo{journal}{Ann. Phys. (Leipzig)} \textbf{\bibinfo{volume}{9}},
  \bibinfo{pages}{885} (\bibinfo{year}{2000}).

\bibitem[{\citenamefont{Aleiner et~al.}(2002)\citenamefont{Aleiner, Brouwer,
  and Glazman}}]{Aleiner2002}
\bibinfo{author}{\bibfnamefont{I.~L.} \bibnamefont{Aleiner}},
  \bibinfo{author}{\bibfnamefont{P.~W.} \bibnamefont{Brouwer}},
  \bibnamefont{and} \bibinfo{author}{\bibfnamefont{L.~I.}
  \bibnamefont{Glazman}}, \bibinfo{journal}{Phys. Rep.}
  \textbf{\bibinfo{volume}{358}}, \bibinfo{pages}{309} (\bibinfo{year}{2002}).

\bibitem[{\citenamefont{Fujisawa et~al.}(2006)\citenamefont{Fujisawa, Hayashi,
  and Sasaki}}]{Fujisawa2006}
\bibinfo{author}{\bibfnamefont{T.}~\bibnamefont{Fujisawa}},
  \bibinfo{author}{\bibfnamefont{T.}~\bibnamefont{Hayashi}}, \bibnamefont{and}
  \bibinfo{author}{\bibfnamefont{S.}~\bibnamefont{Sasaki}},
  \bibinfo{journal}{Rep. Prog. Phys.} \textbf{\bibinfo{volume}{69}},
  \bibinfo{pages}{759} (\bibinfo{year}{2006}).

\bibitem[{\citenamefont{Cuevas and Scheer}(2010)}]{cuevasscheer2010}
\bibinfo{author}{\bibfnamefont{J.~C.} \bibnamefont{Cuevas}} \bibnamefont{and}
  \bibinfo{author}{\bibfnamefont{E.}~\bibnamefont{Scheer}},
  \emph{\bibinfo{title}{Molecular Electronics: An Introduction To Theory And
  Experiment}} (\bibinfo{publisher}{World Scientific},
  \bibinfo{address}{Singapore}, \bibinfo{year}{2010}).

\bibitem[{\citenamefont{Baldea}(2015)}]{Baldea15}
\bibinfo{author}{\bibfnamefont{I.}~\bibnamefont{Baldea}},
  \emph{\bibinfo{title}{Molecular Electronics - An Experimental and Theoretical
  Appraoch}} (\bibinfo{publisher}{Pan Stanford Publishing},
  \bibinfo{address}{Singapore}, \bibinfo{year}{2015}).

\bibitem[{\citenamefont{Datta}(1997)}]{datta1997electronic}
\bibinfo{author}{\bibfnamefont{S.}~\bibnamefont{Datta}},
  \emph{\bibinfo{title}{{Electronic Transport in Mesoscopic Systems}}} (\bibinfo{publisher}{Cambridge
  University Press}, \bibinfo{year}{1997}).

\bibitem[{\citenamefont{Andergassen et~al.}(2010)\citenamefont{Andergassen,
  Meden, Schoeller, Splettstoesser, and Wegewijs}}]{Andergassen2010}
\bibinfo{author}{\bibfnamefont{S.}~\bibnamefont{Andergassen}},
  \bibinfo{author}{\bibfnamefont{V.}~\bibnamefont{Meden}},
  \bibinfo{author}{\bibfnamefont{H.}~\bibnamefont{Schoeller}},
  \bibinfo{author}{\bibfnamefont{J.}~\bibnamefont{Splettstoesser}},
  \bibnamefont{and} \bibinfo{author}{\bibfnamefont{M.}~\bibnamefont{Wegewijs}},
  \bibinfo{journal}{Nanotechnology} \textbf{\bibinfo{volume}{21}},
  \bibinfo{pages}{272001} (\bibinfo{year}{2010}).

\bibitem[{\citenamefont{Haldane}(1981)}]{Haldane1981}
\bibinfo{author}{\bibfnamefont{F.~D.~M.} \bibnamefont{Haldane}},
  \bibinfo{journal}{J. Phys. C}
  \textbf{\bibinfo{volume}{14}}, \bibinfo{pages}{2585} (\bibinfo{year}{1981}).

\bibitem[{\citenamefont{Giamarchi}(2003)}]{giamarchi2003quantum}
\bibinfo{author}{\bibfnamefont{T.}~\bibnamefont{Giamarchi}},
  \emph{\bibinfo{title}{{Quantum Physics in One Dimension}}} (\bibinfo{publisher}{Clarendon Press},
  \bibinfo{year}{2003}).

\bibitem[{\citenamefont{Kane and Fisher}(1992{\natexlab{a}})}]{Kane1992}
\bibinfo{author}{\bibfnamefont{C.~L.} \bibnamefont{Kane}} \bibnamefont{and}
  \bibinfo{author}{\bibfnamefont{M.~P.~A.} \bibnamefont{Fisher}},
  \bibinfo{journal}{Phys. Rev. Lett.} \textbf{\bibinfo{volume}{68}},
  \bibinfo{pages}{1220} (\bibinfo{year}{1992}{\natexlab{a}}).

\bibitem[{\citenamefont{Kane and Fisher}(1992{\natexlab{b}})}]{Kane1992a}
\bibinfo{author}{\bibfnamefont{C.~L.}~\bibnamefont{Kane}} \bibnamefont{and}
  \bibinfo{author}{\bibfnamefont{M.~P.~A.}~\bibnamefont{Fisher}},
  \bibinfo{journal}{Phys. Rev. B} \textbf{\bibinfo{volume}{46}},
  \bibinfo{pages}{15233} (\bibinfo{year}{1992}{\natexlab{b}}).

\bibitem[{\citenamefont{Auslaender et~al.}(2000)\citenamefont{Auslaender,
  Yacoby, {de Picciotto}, Baldwin, Pfeiffer, and West}}]{Auslaender2000}
\bibinfo{author}{\bibfnamefont{O.~M.} \bibnamefont{Auslaender}},
  \bibinfo{author}{\bibfnamefont{A.}~\bibnamefont{Yacoby}},
  \bibinfo{author}{\bibfnamefont{R.}~\bibnamefont{{de Picciotto}}},
  \bibinfo{author}{\bibfnamefont{K.~W.} \bibnamefont{Baldwin}},
  \bibinfo{author}{\bibfnamefont{L.~N.} \bibnamefont{Pfeiffer}},
  \bibnamefont{and} \bibinfo{author}{\bibfnamefont{K.~W.} \bibnamefont{West}},
  \bibinfo{journal}{Phys. Rev. Lett.} \textbf{\bibinfo{volume}{84}},
  \bibinfo{pages}{1764} (\bibinfo{year}{2000}).

\bibitem[{\citenamefont{Sch\"afer et~al.}(2009)\citenamefont{Sch\"afer, Meyer,
  Blumenstein, Roensch, Claessen, Mietke, Klinke, Podlich, Matzdorf,
  Stekolnikov et~al.}}]{Matzdorf2009}
\bibinfo{author}{\bibfnamefont{J.}~\bibnamefont{Sch\"afer}},
  \bibinfo{author}{\bibfnamefont{S.}~\bibnamefont{Meyer}},
  \bibinfo{author}{\bibfnamefont{C.}~\bibnamefont{Blumenstein}},
  \bibinfo{author}{\bibfnamefont{K.}~\bibnamefont{Roensch}},
  \bibinfo{author}{\bibfnamefont{R.}~\bibnamefont{Claessen}},
  \bibinfo{author}{\bibfnamefont{S.}~\bibnamefont{Mietke}},
  \bibinfo{author}{\bibfnamefont{M.}~\bibnamefont{Klinke}},
  \bibinfo{author}{\bibfnamefont{T.}~\bibnamefont{Podlich}},
  \bibinfo{author}{\bibfnamefont{R.}~\bibnamefont{Matzdorf}},
  \bibinfo{author}{\bibfnamefont{A.~A.} \bibnamefont{Stekolnikov}},
  \bibnamefont{et~al.}, \bibinfo{journal}{New J. Phys.}
  \textbf{\bibinfo{volume}{11}}, \bibinfo{pages}{125011}
  (\bibinfo{year}{2009}).

\bibitem[{\citenamefont{Blumenstein et~al.}(2011)\citenamefont{Blumenstein,
  Sch\"afer, Mietke, Meyer, Dollinger, Lochner, Cui, Patthey, Matzdorf, and
  Claessen}}]{Matzdorf2011}
\bibinfo{author}{\bibfnamefont{C.}~\bibnamefont{Blumenstein}},
  \bibinfo{author}{\bibfnamefont{J.}~\bibnamefont{Sch\"afer}},
  \bibinfo{author}{\bibfnamefont{S.}~\bibnamefont{Mietke}},
  \bibinfo{author}{\bibfnamefont{S.}~\bibnamefont{Meyer}},
  \bibinfo{author}{\bibfnamefont{A.}~\bibnamefont{Dollinger}},
  \bibinfo{author}{\bibfnamefont{M.}~\bibnamefont{Lochner}},
  \bibinfo{author}{\bibfnamefont{X.~Y.} \bibnamefont{Cui}},
  \bibinfo{author}{\bibfnamefont{L.}~\bibnamefont{Patthey}},
  \bibinfo{author}{\bibfnamefont{R.}~\bibnamefont{Matzdorf}}, \bibnamefont{and}
  \bibinfo{author}{\bibfnamefont{R.}~\bibnamefont{Claessen}},
  \bibinfo{journal}{Nature Physics} \textbf{\bibinfo{volume}{7}},
  \bibinfo{pages}{776} (\bibinfo{year}{2011}).

\bibitem[{\citenamefont{Laroche et~al.}(2014)\citenamefont{Laroche, Gervais,
  Lilly, and Reno}}]{Laroche2014}
\bibinfo{author}{\bibfnamefont{D.}~\bibnamefont{Laroche}},
  \bibinfo{author}{\bibfnamefont{G.}~\bibnamefont{Gervais}},
  \bibinfo{author}{\bibfnamefont{M.~P.} \bibnamefont{Lilly}}, \bibnamefont{and}
  \bibinfo{author}{\bibfnamefont{J.~L.} \bibnamefont{Reno}},
  \bibinfo{journal}{Science} \textbf{\bibinfo{volume}{343}},
  \bibinfo{pages}{631} (\bibinfo{year}{2014}).

\bibitem[{\citenamefont{Milliken et~al.}(1994)\citenamefont{Milliken, Umbach,
  and Webb}}]{Milliken1994}
\bibinfo{author}{\bibfnamefont{F.~P.} \bibnamefont{Milliken}},
  \bibinfo{author}{\bibfnamefont{C.~P.} \bibnamefont{Umbach}},
  \bibnamefont{and} \bibinfo{author}{\bibfnamefont{R.~A.} \bibnamefont{Webb}},
  \bibinfo{journal}{IBM preprint}  (\bibinfo{year}{1994}).

\bibitem[{\citenamefont{Chang}(2003)}]{Chang2003}
\bibinfo{author}{\bibfnamefont{A.~M.} \bibnamefont{Chang}},
  \bibinfo{journal}{Rev. Mod. Phys.} \textbf{\bibinfo{volume}{75}},
  \bibinfo{pages}{1449} (\bibinfo{year}{2003}).

\bibitem[{\citenamefont{Wurstbauer et~al.}(2013)\citenamefont{Wurstbauer, West,
  Pfeiffer, and Pinczuk}}]{Wurstbauer2013}
\bibinfo{author}{\bibfnamefont{U.}~\bibnamefont{Wurstbauer}},
  \bibinfo{author}{\bibfnamefont{K.~W.} \bibnamefont{West}},
  \bibinfo{author}{\bibfnamefont{L.~N.} \bibnamefont{Pfeiffer}},
  \bibnamefont{and} \bibinfo{author}{\bibfnamefont{A.}~\bibnamefont{Pinczuk}},
  \bibinfo{journal}{Phys. Rev. Lett.} \textbf{\bibinfo{volume}{110}},
  \bibinfo{pages}{026801} (\bibinfo{year}{2013}).

\bibitem[{\citenamefont{Pascher et~al.}(2014)\citenamefont{Pascher, R\"ossler,
  Ihn, Ensslin, Reichl, and Wegscheider}}]{Pascher2014}
\bibinfo{author}{\bibfnamefont{N.}~\bibnamefont{Pascher}},
  \bibinfo{author}{\bibfnamefont{C.}~\bibnamefont{R\"ossler}},
  \bibinfo{author}{\bibfnamefont{T.}~\bibnamefont{Ihn}},
  \bibinfo{author}{\bibfnamefont{K.}~\bibnamefont{Ensslin}},
  \bibinfo{author}{\bibfnamefont{C.}~\bibnamefont{Reichl}}, \bibnamefont{and}
  \bibinfo{author}{\bibfnamefont{W.}~\bibnamefont{Wegscheider}},
  \bibinfo{journal}{Phys. Rev. X} \textbf{\bibinfo{volume}{4}},
  \bibinfo{pages}{011014} (\bibinfo{year}{2014}).

\bibitem[{\citenamefont{Li et~al.}(2015)\citenamefont{Li, Wang, Fu, Du,
  Schreiber, Mu, Liu, Sullivan, Cs\'athy, Lin et~al.}}]{Li2015}
\bibinfo{author}{\bibfnamefont{T.}~\bibnamefont{Li}},
  \bibinfo{author}{\bibfnamefont{P.}~\bibnamefont{Wang}},
  \bibinfo{author}{\bibfnamefont{H.}~\bibnamefont{Fu}},
  \bibinfo{author}{\bibfnamefont{L.}~\bibnamefont{Du}},
  \bibinfo{author}{\bibfnamefont{K.~A.} \bibnamefont{Schreiber}},
  \bibinfo{author}{\bibfnamefont{X.}~\bibnamefont{Mu}},
  \bibinfo{author}{\bibfnamefont{X.}~\bibnamefont{Liu}},
  \bibinfo{author}{\bibfnamefont{G.}~\bibnamefont{Sullivan}},
  \bibinfo{author}{\bibfnamefont{G.~A.} \bibnamefont{Cs\'athy}},
  \bibinfo{author}{\bibfnamefont{X.}~\bibnamefont{Lin}}, \bibnamefont{et~al.},
  \bibinfo{journal}{Phys.\ Rev.\ Lett.\,} \textbf{\bibinfo{volume}{115}},
  \bibinfo{pages}{136804} (\bibinfo{year}{2015}).

\bibitem[{\citenamefont{{von Delft} and Schoeller}(1998)}]{vonDelft1998}
\bibinfo{author}{\bibfnamefont{J.}~\bibnamefont{{von Delft}}} \bibnamefont{and}
  \bibinfo{author}{\bibfnamefont{H.}~\bibnamefont{Schoeller}},
  \bibinfo{journal}{Ann. Phys.} \textbf{\bibinfo{volume}{7}},
  \bibinfo{pages}{225} (\bibinfo{year}{1998}).

\bibitem[{\citenamefont{Sch\"onhammer}(2005)}]{Schoenhammer2005}
\bibinfo{author}{\bibfnamefont{K.}~\bibnamefont{Sch\"onhammer}},
  \emph{\bibinfo{title}{Interacting Electrons in Low Dimensions}}
  (\bibinfo{publisher}{Springer}, \bibinfo{address}{Berlin, Germany},
  \bibinfo{year}{2005}). \eprint{arXiv:cond-mat/0305035}.

\bibitem[{\citenamefont{Metzner et~al.}(2012)\citenamefont{Metzner, Salmhofer,
  Honerkamp, Meden, and Sch\"onhammer}}]{Metzner2012}
\bibinfo{author}{\bibfnamefont{W.}~\bibnamefont{Metzner}},
  \bibinfo{author}{\bibfnamefont{M.}~\bibnamefont{Salmhofer}},
  \bibinfo{author}{\bibfnamefont{C.}~\bibnamefont{Honerkamp}},
  \bibinfo{author}{\bibfnamefont{V.}~\bibnamefont{Meden}}, \bibnamefont{and}
  \bibinfo{author}{\bibfnamefont{K.}~\bibnamefont{Sch\"onhammer}},
  \bibinfo{journal}{Rev. Mod. Phys.} \textbf{\bibinfo{volume}{84}},
  \bibinfo{pages}{299} (\bibinfo{year}{2012}).

\bibitem[{\citenamefont{Furusaki and Nagaosa}(1993)}]{Furusaki1993}
\bibinfo{author}{\bibfnamefont{A.}~\bibnamefont{Furusaki}} \bibnamefont{and}
  \bibinfo{author}{\bibfnamefont{N.}~\bibnamefont{Nagaosa}},
  \bibinfo{journal}{Phys. Rev. B} \textbf{\bibinfo{volume}{47}},
  \bibinfo{pages}{4631} (\bibinfo{year}{1993}).

\bibitem[{\citenamefont{Matveev et~al.}(1993)\citenamefont{Matveev, Yue, and
  Glazman}}]{Matveev1993}
\bibinfo{author}{\bibfnamefont{K.~A.} \bibnamefont{Matveev}},
  \bibinfo{author}{\bibfnamefont{D.}~\bibnamefont{Yue}}, \bibnamefont{and}
  \bibinfo{author}{\bibfnamefont{L.~I.} \bibnamefont{Glazman}},
  \bibinfo{journal}{Phys. Rev. Lett.} \textbf{\bibinfo{volume}{71}},
  \bibinfo{pages}{3351} (\bibinfo{year}{1993}).

\bibitem[{\citenamefont{Yue et~al.}(1994)\citenamefont{Yue, Glazman, and
  Matveev}}]{Yue1994}
\bibinfo{author}{\bibfnamefont{D.}~\bibnamefont{Yue}},
  \bibinfo{author}{\bibfnamefont{L.~I.} \bibnamefont{Glazman}},
  \bibnamefont{and} \bibinfo{author}{\bibfnamefont{K.~A.}
  \bibnamefont{Matveev}}, \bibinfo{journal}{Phys. Rev. B}
  \textbf{\bibinfo{volume}{49}}, \bibinfo{pages}{1966} (\bibinfo{year}{1994}).

\bibitem[{\citenamefont{Moon et~al.}(1993)\citenamefont{Moon, Yi, Kane, Girvin,
  and Fisher}}]{Moon1995}
\bibinfo{author}{\bibfnamefont{K.}~\bibnamefont{Moon}},
  \bibinfo{author}{\bibfnamefont{H.}~\bibnamefont{Yi}},
  \bibinfo{author}{\bibfnamefont{C.~L.} \bibnamefont{Kane}},
  \bibinfo{author}{\bibfnamefont{S.~M.} \bibnamefont{Girvin}},
  \bibnamefont{and} \bibinfo{author}{\bibfnamefont{M.~P.~A.}
  \bibnamefont{Fisher}}, \bibinfo{journal}{Phys. Rev. Lett.}
  \textbf{\bibinfo{volume}{71}}, \bibinfo{pages}{4381} (\bibinfo{year}{1993}).

\bibitem[{\citenamefont{Egger and Grabert}(1995)}]{Egger1995}
\bibinfo{author}{\bibfnamefont{R.}~\bibnamefont{Egger}} \bibnamefont{and}
  \bibinfo{author}{\bibfnamefont{H.}~\bibnamefont{Grabert}},
  \bibinfo{journal}{Phys. Rev. Lett.} \textbf{\bibinfo{volume}{75}},
  \bibinfo{pages}{3505} (\bibinfo{year}{1995}).

\bibitem[{\citenamefont{Fendley et~al.}(1995)\citenamefont{Fendley, Ludwig, and
  Saleur}}]{Fendley1995}
\bibinfo{author}{\bibfnamefont{P.}~\bibnamefont{Fendley}},
  \bibinfo{author}{\bibfnamefont{A.~W.~W.}~\bibnamefont{Ludwig}}, \bibnamefont{and}
  \bibinfo{author}{\bibfnamefont{H.}~\bibnamefont{Saleur}},
  \bibinfo{journal}{Phys. Rev. Lett.} \textbf{\bibinfo{volume}{74}},
  \bibinfo{pages}{3005} (\bibinfo{year}{1995}).

\bibitem[{\citenamefont{Furusaki}(1997)}]{Furusaki1997}
\bibinfo{author}{\bibfnamefont{A.}~\bibnamefont{Furusaki}},
  \bibinfo{journal}{Phys. Rev. B} \textbf{\bibinfo{volume}{56}},
  \bibinfo{pages}{9352} (\bibinfo{year}{1997}).

\bibitem[{\citenamefont{Aristov and W\"olfle}(2008)}]{Aristov2008}
\bibinfo{author}{\bibfnamefont{D.~N.} \bibnamefont{Aristov}} \bibnamefont{and}
  \bibinfo{author}{\bibfnamefont{P.}~\bibnamefont{W\"olfle}},
  \bibinfo{journal}{Europhys. Lett.} \textbf{\bibinfo{volume}{82}},
  \bibinfo{pages}{27001} (\bibinfo{year}{2008}).

\bibitem[{\citenamefont{Aristov and W\"olfle}(2009)}]{Aristov2009}
\bibinfo{author}{\bibfnamefont{D.~N.} \bibnamefont{Aristov}} \bibnamefont{and}
  \bibinfo{author}{\bibfnamefont{P.}~\bibnamefont{W\"olfle}},
  \bibinfo{journal}{Phys. Rev. B} \textbf{\bibinfo{volume}{80}},
  \bibinfo{pages}{045109} (\bibinfo{year}{2009}).

\bibitem[{\citenamefont{Einhellinger et~al.}(2012)\citenamefont{Einhellinger,
  Cojuhovschi, and Jeckelmann}}]{Einhellinger2012}
\bibinfo{author}{\bibfnamefont{M.}~\bibnamefont{Einhellinger}},
  \bibinfo{author}{\bibfnamefont{A.}~\bibnamefont{Cojuhovschi}},
  \bibnamefont{and}
  \bibinfo{author}{\bibfnamefont{E.}~\bibnamefont{Jeckelmann}},
  \bibinfo{journal}{Phys. Rev. B} \textbf{\bibinfo{volume}{85}},
  \bibinfo{pages}{235141} (\bibinfo{year}{2012}).

\bibitem[{\citenamefont{Fabrizio et~al.}(1994)\citenamefont{Fabrizio, Gogolin,
  and Scheidl}}]{Fabrizio1994}
\bibinfo{author}{\bibfnamefont{M.}~\bibnamefont{Fabrizio}},
  \bibinfo{author}{\bibfnamefont{A.~O.} \bibnamefont{Gogolin}},
  \bibnamefont{and} \bibinfo{author}{\bibfnamefont{S.}~\bibnamefont{Scheidl}},
  \bibinfo{journal}{Phys. Rev. Lett.} \textbf{\bibinfo{volume}{72}},
  \bibinfo{pages}{2235} (\bibinfo{year}{1994}).

\bibitem[{\citenamefont{Maurey and Giamarchi}(1995)}]{Maurey1995}
\bibinfo{author}{\bibfnamefont{H.}~\bibnamefont{Maurey}} \bibnamefont{and}
  \bibinfo{author}{\bibfnamefont{T.}~\bibnamefont{Giamarchi}},
  \bibinfo{journal}{Phys. Rev. B} \textbf{\bibinfo{volume}{51}},
  \bibinfo{pages}{10833} (\bibinfo{year}{1995}).

\bibitem[{\citenamefont{Weiss et~al.}(1995)\citenamefont{Weiss, Egger, and
  Sassetti}}]{Weiss1995}
\bibinfo{author}{\bibfnamefont{U.}~\bibnamefont{Weiss}},
  \bibinfo{author}{\bibfnamefont{R.}~\bibnamefont{Egger}}, \bibnamefont{and}
  \bibinfo{author}{\bibfnamefont{M.}~\bibnamefont{Sassetti}},
  \bibinfo{journal}{Phys. Rev. B} \textbf{\bibinfo{volume}{52}}, \bibinfo{pages}{16707}
  (\bibinfo{year}{1995}).

\bibitem[{\citenamefont{Nazarov and Glazman}(2003)}]{Nazarov2003}
\bibinfo{author}{\bibfnamefont{Y.~V.} \bibnamefont{Nazarov}} \bibnamefont{and}
  \bibinfo{author}{\bibfnamefont{L.~I.} \bibnamefont{Glazman}},
  \bibinfo{journal}{Phys. Rev. Lett.} \textbf{\bibinfo{volume}{91}},
  \bibinfo{pages}{126804} (\bibinfo{year}{2003}).

\bibitem[{\citenamefont{Komnik and Gogolin}(2003{\natexlab{a}})}]{Komnik2003}
\bibinfo{author}{\bibfnamefont{A.}~\bibnamefont{Komnik}} \bibnamefont{and}
  \bibinfo{author}{\bibfnamefont{A.~O.}~\bibnamefont{Gogolin}},
  \bibinfo{journal}{Phys. Rev. Lett.} \textbf{\bibinfo{volume}{90}},
  \bibinfo{pages}{246403} (\bibinfo{year}{2003}{\natexlab{a}}).

\bibitem[{\citenamefont{Komnik and Gogolin}(2003{\natexlab{b}})}]{Komnik2003a}
\bibinfo{author}{\bibfnamefont{A.}~\bibnamefont{Komnik}} \bibnamefont{and}
  \bibinfo{author}{\bibfnamefont{A.~O.}~\bibnamefont{Gogolin}},
  \bibinfo{journal}{Phys. Rev. B} \textbf{\bibinfo{volume}{68}},
  \bibinfo{pages}{235323} (\bibinfo{year}{2003}{\natexlab{b}}).

\bibitem[{\citenamefont{W\"achter et~al.}(2007)\citenamefont{W\"achter, Meden,
  and Sch\"onhammer}}]{Waechter2007}
\bibinfo{author}{\bibfnamefont{P.}~\bibnamefont{W\"achter}},
  \bibinfo{author}{\bibfnamefont{V.}~\bibnamefont{Meden}}, \bibnamefont{and}
  \bibinfo{author}{\bibfnamefont{K.}~\bibnamefont{Sch\"onhammer}},
  \bibinfo{journal}{Phys. Rev. B} \textbf{\bibinfo{volume}{76}},
  \bibinfo{pages}{125316} (\bibinfo{year}{2007}).

\bibitem[{\citenamefont{Goldstein et~al.}(2010)\citenamefont{Goldstein, Weiss,
  and Berkovits}}]{Goldstein2010}
\bibinfo{author}{\bibfnamefont{M.}~\bibnamefont{Goldstein}},
  \bibinfo{author}{\bibfnamefont{Y.}~\bibnamefont{Weiss}}, \bibnamefont{and}
  \bibinfo{author}{\bibfnamefont{R.}~\bibnamefont{Berkovits}},
  \bibinfo{journal}{Physica E} \textbf{\bibinfo{volume}{42}},
  \bibinfo{pages}{610} (\bibinfo{year}{2010}).

\bibitem[{\citenamefont{Goldstein and
  Berkovits}(2010{\natexlab{a}})}]{Goldstein2010b}
\bibinfo{author}{\bibfnamefont{M.}~\bibnamefont{Goldstein}} \bibnamefont{and}
  \bibinfo{author}{\bibfnamefont{R.}~\bibnamefont{Berkovits}},
  \bibinfo{journal}{Phys. Rev. B} \textbf{\bibinfo{volume}{82}},
  \bibinfo{pages}{235315} (\bibinfo{year}{2010}{\natexlab{a}}).

\bibitem[{\citenamefont{Lerner et~al.}(2008)\citenamefont{Lerner, Yudson, and
  Yurkevich}}]{Lerner2008}
\bibinfo{author}{\bibfnamefont{I.~V.} \bibnamefont{Lerner}},
  \bibinfo{author}{\bibfnamefont{V.~I.} \bibnamefont{Yudson}},
  \bibnamefont{and} \bibinfo{author}{\bibfnamefont{I.~V.}
  \bibnamefont{Yurkevich}}, \bibinfo{journal}{Phys. Rev. Lett.}
  \textbf{\bibinfo{volume}{100}}, \bibinfo{pages}{256805} (\bibinfo{year}{2008}).

\bibitem[{\citenamefont{Goldstein and
  Berkovits}(2010{\natexlab{b}})}]{Goldstein2010a}
\bibinfo{author}{\bibfnamefont{M.}~\bibnamefont{Goldstein}} \bibnamefont{and}
  \bibinfo{author}{\bibfnamefont{R.}~\bibnamefont{Berkovits}},
  \bibinfo{journal}{Phys. Rev. Lett.} \textbf{\bibinfo{volume}{104}},
  \bibinfo{pages}{106403} (\bibinfo{year}{2010}{\natexlab{b}}).

\bibitem[{\citenamefont{Andergassen et~al.}(2006)\citenamefont{Andergassen,
  Enss, and Meden}}]{Andergassen2006}
\bibinfo{author}{\bibfnamefont{S.}~\bibnamefont{Andergassen}},
  \bibinfo{author}{\bibfnamefont{T.}~\bibnamefont{Enss}}, \bibnamefont{and}
  \bibinfo{author}{\bibfnamefont{V.}~\bibnamefont{Meden}},
  \bibinfo{journal}{Phys. Rev. B} \textbf{\bibinfo{volume}{73}},
  \bibinfo{pages}{153308} (\bibinfo{year}{2006}).

\bibitem[{\citenamefont{Jeckelmann}(2012)}]{Jeckelmann2012}
\bibinfo{author}{\bibfnamefont{E.}~\bibnamefont{Jeckelmann}},
  \bibinfo{journal}{J. Phys. Condens. Matter} \textbf{\bibinfo{volume}{25}},
  \bibinfo{pages}{014002} (\bibinfo{year}{2012}).

\bibitem[{\citenamefont{Elste et~al.}(2011{\natexlab{a}})\citenamefont{Elste,
  Reichman, and Millis}}]{Elste2011a}
\bibinfo{author}{\bibfnamefont{F.}~\bibnamefont{Elste}},
  \bibinfo{author}{\bibfnamefont{D.~R.}~\bibnamefont{Reichman}}, \bibnamefont{and}
  \bibinfo{author}{\bibfnamefont{A.~J.}~\bibnamefont{Millis}},
  \bibinfo{journal}{Phys. Rev. B} \textbf{\bibinfo{volume}{83}},
  \bibinfo{pages}{085415} (\bibinfo{year}{2011}{\natexlab{a}}).

\bibitem[{\citenamefont{Matveev}(1992)}]{Matveev1992}
\bibinfo{author}{\bibfnamefont{K.~A.}~\bibnamefont{Matveev}} \bibnamefont{and}
\bibinfo{author}{\bibfnamefont{A.~I.}~\bibnamefont{Larkin}} 
  \bibinfo{journal}{Phys. Rev. B} \textbf{\bibinfo{volume}{46}},
  \bibinfo{pages}{15337} (\bibinfo{year}{1992}).  
  
\bibitem[{\citenamefont{Dubi}(2013)}]{Dubi2013}
\bibinfo{author}{\bibfnamefont{Y.}~\bibnamefont{Dubi}}, \bibinfo{journal}{J.
  Chem. Phys.} \textbf{\bibinfo{volume}{139}}, \bibinfo{pages}{154710}
  (\bibinfo{year}{2013}).
  
\bibitem[{\citenamefont{Furusaki}(1998)}]{Furusaki1998}
\bibinfo{author}{\bibfnamefont{A.}~\bibnamefont{Furusaki}},
  \bibinfo{journal}{Phys. Rev. B} \textbf{\bibinfo{volume}{57}},
  \bibinfo{pages}{7141} (\bibinfo{year}{1998}).

\bibitem[{\citenamefont{Durganandini}(2006{\natexlab{a}})}]{Durganandini2006}
\bibinfo{author}{\bibfnamefont{P.}~\bibnamefont{Durganandini}}, \bibinfo{journal}{Phys. Rev. B}
  \textbf{\bibinfo{volume}{73}}, \bibinfo{pages}{115316}
  (\bibinfo{year}{2006}{\natexlab{a}}).

\bibitem[{\citenamefont{Durganandini}(2006{\natexlab{b}})}]{Durganandini2006b}
\bibinfo{author}{\bibfnamefont{P.}~\bibnamefont{Durganandini}},
  \bibinfo{journal}{Phys. Rev. B} \textbf{\bibinfo{volume}{74}},
  \bibinfo{pages}{155309} (\bibinfo{year}{2006}{\natexlab{b}}).

\bibitem[{\citenamefont{Kawaguchi}(2009)}]{Kawaguchi2009}
\bibinfo{author}{\bibfnamefont{S.}~\bibnamefont{Kawaguchi}},
  \bibinfo{journal}{J. Phys. Condens. Matt.} \textbf{\bibinfo{volume}{21}},
  \bibinfo{pages}{395303} (\bibinfo{year}{2009}).

\bibitem[{\citenamefont{Yang et~al.}(2014)\citenamefont{Yang, He, Wang, Liu,
  and Liu}}]{Yang2014}
\bibinfo{author}{\bibfnamefont{K.-H.}~\bibnamefont{Yang}},
  \bibinfo{author}{\bibfnamefont{X.}~\bibnamefont{He}},
  \bibinfo{author}{\bibfnamefont{H.-Y.} \bibnamefont{Wang}},
  \bibinfo{author}{\bibfnamefont{K.-D.} \bibnamefont{Liu}}, \bibnamefont{and}
  \bibinfo{author}{\bibfnamefont{B.-Y.} \bibnamefont{Liu}},
  \bibinfo{journal}{Phys. Lett. A} \textbf{\bibinfo{volume}{378}},
  \bibinfo{pages}{3136} (\bibinfo{year}{2014}).

\bibitem[{\citenamefont{Maier and Komnik}(2010)}]{Maier2010}
\bibinfo{author}{\bibfnamefont{S.}~\bibnamefont{Maier}} \bibnamefont{and}
  \bibinfo{author}{\bibfnamefont{A.}~\bibnamefont{Komnik}},
  \bibinfo{journal}{Phys. Rev. B} \textbf{\bibinfo{volume}{82}},
  \bibinfo{pages}{165116} (\bibinfo{year}{2010}).

\bibitem[{\citenamefont{Yang and Zhao}(2010)}]{Yang2010}
\bibinfo{author}{\bibfnamefont{K.-H.}~\bibnamefont{Yang}} \bibnamefont{and}
  \bibinfo{author}{\bibfnamefont{Y.-L.}~\bibnamefont{Zhao}}, \bibinfo{journal}{Physica E}
  \textbf{\bibinfo{volume}{42}}, \bibinfo{pages}{2324} (\bibinfo{year}{2010}).

\bibitem[{\citenamefont{Skorobagatko}(2012)}]{Skorobagatko2012}
\bibinfo{author}{\bibfnamefont{G.~A.}~\bibnamefont{Skorobagatko}}, \bibinfo{journal}{Phys. Rev. B}
  \textbf{\bibinfo{volume}{85}}, \bibinfo{pages}{075310}
  (\bibinfo{year}{2012}).

\bibitem[{\citenamefont{Hattori and Rosch}(2014)}]{Hattori2014}
\bibinfo{author}{\bibfnamefont{K.}~\bibnamefont{Hattori}} \bibnamefont{and}
  \bibinfo{author}{\bibfnamefont{A.}~\bibnamefont{Rosch}},
  \bibinfo{journal}{Phys. Rev. B} \textbf{\bibinfo{volume}{90}},
  \bibinfo{pages}{115103} (\bibinfo{year}{2014}).

\bibitem[{\citenamefont{Al-Hassanieh et~al.}(2006)\citenamefont{Al-Hassanieh,
  Feiguin, Riera, B\"usser, and Dagotto}}]{Hassanieh2006}
\bibinfo{author}{\bibfnamefont{K.~A.} \bibnamefont{Al-Hassanieh}},
  \bibinfo{author}{\bibfnamefont{A.~E.} \bibnamefont{Feiguin}},
  \bibinfo{author}{\bibfnamefont{J.~A.} \bibnamefont{Riera}},
  \bibinfo{author}{\bibfnamefont{C.~A.} \bibnamefont{B\"usser}},
  \bibnamefont{and} \bibinfo{author}{\bibfnamefont{E.}~\bibnamefont{Dagotto}},
  \bibinfo{journal}{Phys. Rev. B} \textbf{\bibinfo{volume}{73}},
  \bibinfo{pages}{195304} (\bibinfo{year}{2006}).

\bibitem[{\citenamefont{{Dias da Silva} et~al.}(2008)\citenamefont{{Dias da
  Silva}, {Heidrich-Meisner}, Feiguin, B\"usser, Martins, Anda, and
  Dagotto}}]{daSilva2008}
\bibinfo{author}{\bibfnamefont{L.~G.~G.~V.} \bibnamefont{{Dias~da~Silva}}},
  \bibinfo{author}{\bibfnamefont{F.}~\bibnamefont{{Heidrich-Meisner}}},
  \bibinfo{author}{\bibfnamefont{A.~E.} \bibnamefont{Feiguin}},
  \bibinfo{author}{\bibfnamefont{C.~A.} \bibnamefont{B\"usser}},
  \bibinfo{author}{\bibfnamefont{G.~B.} \bibnamefont{Martins}},
  \bibinfo{author}{\bibfnamefont{E.~V.} \bibnamefont{Anda}}, \bibnamefont{and}
  \bibinfo{author}{\bibfnamefont{E.}~\bibnamefont{Dagotto}},
  \bibinfo{journal}{Phys. Rev. B} \textbf{\bibinfo{volume}{78}},
  \bibinfo{pages}{195317} (\bibinfo{year}{2008}).

\bibitem[{\citenamefont{Kirino et~al.}(2008)\citenamefont{Kirino, Fujii, Zhao,
  and Ueda}}]{Kirino2008}
\bibinfo{author}{\bibfnamefont{S.}~\bibnamefont{Kirino}},
  \bibinfo{author}{\bibfnamefont{T.}~\bibnamefont{Fujii}},
  \bibinfo{author}{\bibfnamefont{J.}~\bibnamefont{Zhao}}, \bibnamefont{and}
  \bibinfo{author}{\bibfnamefont{K.}~\bibnamefont{Ueda}}, \bibinfo{journal}{J.
  Phys. Soc. Jpn.} \textbf{\bibinfo{volume}{77}}, \bibinfo{pages}{084704}
  (\bibinfo{year}{2008}).

\bibitem[{\citenamefont{Heidrich-Meisner
  et~al.}(2009)\citenamefont{Heidrich-Meisner, Feiguin, and
  Dagotto}}]{HeidrichMeisner2009}
\bibinfo{author}{\bibfnamefont{F.}~\bibnamefont{Heidrich-Meisner}},
  \bibinfo{author}{\bibfnamefont{A.~E.} \bibnamefont{Feiguin}},
  \bibnamefont{and} \bibinfo{author}{\bibfnamefont{E.}~\bibnamefont{Dagotto}},
  \bibinfo{journal}{Phys. Rev. B} \textbf{\bibinfo{volume}{79}},
  \bibinfo{pages}{235336} (\bibinfo{year}{2009}).

\bibitem[{\citenamefont{Kirino et~al.}(2010)\citenamefont{Kirino, Fujii, and
  Ueda}}]{Kirino2010}
\bibinfo{author}{\bibfnamefont{S.}~\bibnamefont{Kirino}},
  \bibinfo{author}{\bibfnamefont{T.}~\bibnamefont{Fujii}}, \bibnamefont{and}
  \bibinfo{author}{\bibfnamefont{K.}~\bibnamefont{Ueda}},
  \bibinfo{journal}{Physica E} \textbf{\bibinfo{volume}{42}},
  \bibinfo{pages}{874} (\bibinfo{year}{2010}).

\bibitem[{\citenamefont{Bransch\"adel et~al.}(2010)\citenamefont{Bransch\"adel,
  Schneider, and Schmitteckert}}]{Branschadel2010}
\bibinfo{author}{\bibfnamefont{A.}~\bibnamefont{Bransch\"adel}},
  \bibinfo{author}{\bibfnamefont{G.}~\bibnamefont{Schneider}},
  \bibnamefont{and}
  \bibinfo{author}{\bibfnamefont{P.}~\bibnamefont{Schmitteckert}},
  \bibinfo{journal}{Ann. Phys.} \textbf{\bibinfo{volume}{522}},
  \bibinfo{pages}{657} (\bibinfo{year}{2010}).

\bibitem[{\citenamefont{Anders}(2008{\natexlab{a}})}]{Anders2008}
\bibinfo{author}{\bibfnamefont{F.~B.} \bibnamefont{Anders}},
  \bibinfo{journal}{Phys. Rev. Lett.} \textbf{\bibinfo{volume}{101}},
  \bibinfo{pages}{066804} (\bibinfo{year}{2008}{\natexlab{a}}).

\bibitem[{\citenamefont{Anders}(2008{\natexlab{b}})}]{Anders2008b}
\bibinfo{author}{\bibfnamefont{F.~B.} \bibnamefont{Anders}},
  \bibinfo{journal}{J. Phys. Condens. Matter} \textbf{\bibinfo{volume}{20}},
  \bibinfo{pages}{195216} (\bibinfo{year}{2008}{\natexlab{b}}).

\bibitem[{\citenamefont{Schmitt and Anders}(2010)}]{Schmitt2010}
\bibinfo{author}{\bibfnamefont{S.}~\bibnamefont{Schmitt}} \bibnamefont{and}
  \bibinfo{author}{\bibfnamefont{F.~B.}~\bibnamefont{Anders}},
  \bibinfo{journal}{Phys. Rev. B} \textbf{\bibinfo{volume}{81}},
  \bibinfo{pages}{165106} (\bibinfo{year}{2010}).

\bibitem[{\citenamefont{Nghiem and Costi}(2014)}]{Nghiem2014}
\bibinfo{author}{\bibfnamefont{H.~T.~M.}~\bibnamefont{Nghiem}} \bibnamefont{and}
  \bibinfo{author}{\bibfnamefont{T.~A.}~\bibnamefont{Costi}},
  \bibinfo{journal}{Phys. Rev. B} \textbf{\bibinfo{volume}{89}},
  \bibinfo{pages}{075118} (\bibinfo{year}{2014}).
  
\bibitem[{\citenamefont{Wang and Thoss}(2013)}]{Thoss2013}
\bibinfo{author}{\bibfnamefont{H.}~\bibnamefont{Wang}} \bibnamefont{and}
  \bibinfo{author}{\bibfnamefont{M.}~\bibnamefont{Thoss}}, \bibinfo{journal}{J.
  Chem. Phys.} \textbf{\bibinfo{volume}{138}}, \bibinfo{pages}{134704}
  (\bibinfo{year}{2013}).

\bibitem[{\citenamefont{Balzer et~al.}(2015)\citenamefont{Balzer, Li, Vendrell,
  and Eckstein}}]{Balzer2015}
\bibinfo{author}{\bibfnamefont{K.}~\bibnamefont{Balzer}},
  \bibinfo{author}{\bibfnamefont{Z.}~\bibnamefont{Li}},
  \bibinfo{author}{\bibfnamefont{O.}~\bibnamefont{Vendrell}}, \bibnamefont{and}
  \bibinfo{author}{\bibfnamefont{M.}~\bibnamefont{Eckstein}},
  \bibinfo{journal}{Phys. Rev. B} \textbf{\bibinfo{volume}{91}},
  \bibinfo{pages}{045136} (\bibinfo{year}{2015}).

\bibitem[{\citenamefont{Weiss et~al.}(2008)\citenamefont{Weiss, Eckel,
  Thorwart, and Egger}}]{Thorwart2008}
\bibinfo{author}{\bibfnamefont{S.}~\bibnamefont{Weiss}},
  \bibinfo{author}{\bibfnamefont{J.}~\bibnamefont{Eckel}},
  \bibinfo{author}{\bibfnamefont{M.}~\bibnamefont{Thorwart}}, \bibnamefont{and}
  \bibinfo{author}{\bibfnamefont{R.}~\bibnamefont{Egger}},
  \bibinfo{journal}{Phys. Rev. B} \textbf{\bibinfo{volume}{77}},
  \bibinfo{pages}{195316} (\bibinfo{year}{2008}).

\bibitem[{\citenamefont{Segal et~al.}(2010)\citenamefont{Segal, Millis, and
  Reichman}}]{Segal2010}
\bibinfo{author}{\bibfnamefont{D.}~\bibnamefont{Segal}},
  \bibinfo{author}{\bibfnamefont{A.~J.} \bibnamefont{Millis}},
  \bibnamefont{and} \bibinfo{author}{\bibfnamefont{D.~R.}
  \bibnamefont{Reichman}}, \bibinfo{journal}{Phys. Rev. B}
  \textbf{\bibinfo{volume}{82}}, \bibinfo{pages}{205323}
  (\bibinfo{year}{2010}).

\bibitem[{\citenamefont{Segal et~al.}(2011)\citenamefont{Segal, Millis, and
  Reichman}}]{Segal2011}
\bibinfo{author}{\bibfnamefont{D.}~\bibnamefont{Segal}},
  \bibinfo{author}{\bibfnamefont{A.~J.} \bibnamefont{Millis}},
  \bibnamefont{and} \bibinfo{author}{\bibfnamefont{D.~R.}
  \bibnamefont{Reichman}}, \bibinfo{journal}{Phys. Chem. Chem. Phys.}
  \textbf{\bibinfo{volume}{13}}, \bibinfo{pages}{14378} (\bibinfo{year}{2011}).

\bibitem[{\citenamefont{H\"utzen et~al.}(2012)\citenamefont{H\"utzen, Weiss,
  Thorwart, and Egger}}]{Huetzen2012}
\bibinfo{author}{\bibfnamefont{R.}~\bibnamefont{H\"utzen}},
  \bibinfo{author}{\bibfnamefont{S.}~\bibnamefont{Weiss}},
  \bibinfo{author}{\bibfnamefont{M.}~\bibnamefont{Thorwart}}, \bibnamefont{and}
  \bibinfo{author}{\bibfnamefont{R.}~\bibnamefont{Egger}},
  \bibinfo{journal}{Phys. Rev. B} \textbf{\bibinfo{volume}{85}},
  \bibinfo{pages}{121408} (\bibinfo{year}{2012}).

\bibitem[{\citenamefont{Weiss et~al.}(2013)\citenamefont{Weiss, H\"utzen,
  Becker, Eckel, Egger, and Thorwart}}]{Weiss2013}
\bibinfo{author}{\bibfnamefont{S.}~\bibnamefont{Weiss}},
  \bibinfo{author}{\bibfnamefont{R.}~\bibnamefont{H\"utzen}},
  \bibinfo{author}{\bibfnamefont{D.}~\bibnamefont{Becker}},
  \bibinfo{author}{\bibfnamefont{J.}~\bibnamefont{Eckel}},
  \bibinfo{author}{\bibfnamefont{R.}~\bibnamefont{Egger}}, \bibnamefont{and}
  \bibinfo{author}{\bibfnamefont{M.}~\bibnamefont{Thorwart}},
  \bibinfo{journal}{Phys. Status Solidi B} \textbf{\bibinfo{volume}{250}},
  \bibinfo{pages}{2298} (\bibinfo{year}{2013}).

\bibitem[{\citenamefont{Werner et~al.}(2006)\citenamefont{Werner, Comanac, {de'
  Medici}, Troyer, and Millis}}]{Werner2006}
\bibinfo{author}{\bibfnamefont{P.}~\bibnamefont{Werner}},
  \bibinfo{author}{\bibfnamefont{A.}~\bibnamefont{Comanac}},
  \bibinfo{author}{\bibfnamefont{L.}~\bibnamefont{{de' Medici}}},
  \bibinfo{author}{\bibfnamefont{M.}~\bibnamefont{Troyer}}, \bibnamefont{and}
  \bibinfo{author}{\bibfnamefont{A.~J.} \bibnamefont{Millis}},
  \bibinfo{journal}{Phys. Rev. Lett.} \textbf{\bibinfo{volume}{97}},
  \bibinfo{pages}{076405} (\bibinfo{year}{2006}).

\bibitem[{\citenamefont{Schmidt et~al.}(2008)\citenamefont{Schmidt, Werner,
  M\"uhlbacher, and Komnik}}]{Schmidt2008}
\bibinfo{author}{\bibfnamefont{T.~L.} \bibnamefont{Schmidt}},
  \bibinfo{author}{\bibfnamefont{P.}~\bibnamefont{Werner}},
  \bibinfo{author}{\bibfnamefont{L.}~\bibnamefont{M\"uhlbacher}},
  \bibnamefont{and} \bibinfo{author}{\bibfnamefont{A.}~\bibnamefont{Komnik}},
  \bibinfo{journal}{Phys. Rev. B} \textbf{\bibinfo{volume}{78}},
  \bibinfo{pages}{235110} (\bibinfo{year}{2008}).

\bibitem[{\citenamefont{Werner et~al.}(2009)\citenamefont{Werner, Oka, and
  Millis}}]{Werner2009}
\bibinfo{author}{\bibfnamefont{P.}~\bibnamefont{Werner}},
  \bibinfo{author}{\bibfnamefont{T.}~\bibnamefont{Oka}}, \bibnamefont{and}
  \bibinfo{author}{\bibfnamefont{A.~J.} \bibnamefont{Millis}},
  \bibinfo{journal}{Phys. Rev. B} \textbf{\bibinfo{volume}{79}},
  \bibinfo{pages}{035320} (\bibinfo{year}{2009}).

\bibitem[{\citenamefont{Schir\'o}(2010)}]{Schiro2010}
\bibinfo{author}{\bibfnamefont{M.}~\bibnamefont{Schir\'o}},
  \bibinfo{journal}{Phys. Rev. B} \textbf{\bibinfo{volume}{81}},
  \bibinfo{pages}{085126} (\bibinfo{year}{2010}).

\bibitem[{\citenamefont{Gull et~al.}(2011)\citenamefont{Gull, Millis,
  Lichtenstein, Rubtsov, Troyer, and Werner}}]{Gull2010}
\bibinfo{author}{\bibfnamefont{E.}~\bibnamefont{Gull}},
  \bibinfo{author}{\bibfnamefont{A.~J.} \bibnamefont{Millis}},
  \bibinfo{author}{\bibfnamefont{A.~I.} \bibnamefont{Lichtenstein}},
  \bibinfo{author}{\bibfnamefont{A.~N.} \bibnamefont{Rubtsov}},
  \bibinfo{author}{\bibfnamefont{M.}~\bibnamefont{Troyer}}, \bibnamefont{and}
  \bibinfo{author}{\bibfnamefont{P.}~\bibnamefont{Werner}},
  \bibinfo{journal}{Rev. Mod. Phys.} \textbf{\bibinfo{volume}{83}},
  \bibinfo{pages}{349} (\bibinfo{year}{2011}).

\bibitem[{\citenamefont{Cohen and Rabani}(2011)}]{Cohen2011}
\bibinfo{author}{\bibfnamefont{G.}~\bibnamefont{Cohen}} \bibnamefont{and}
  \bibinfo{author}{\bibfnamefont{E.}~\bibnamefont{Rabani}},
  \bibinfo{journal}{Phys. Rev. B} \textbf{\bibinfo{volume}{84}},
  \bibinfo{pages}{075150} (\bibinfo{year}{2011}).

\bibitem[{\citenamefont{M\"uhlbacher et~al.}(2011)\citenamefont{M\"uhlbacher,
  Urban, and Komnik}}]{Muhlbacher2011}
\bibinfo{author}{\bibfnamefont{L.}~\bibnamefont{M\"uhlbacher}},
  \bibinfo{author}{\bibfnamefont{D.~F.} \bibnamefont{Urban}}, \bibnamefont{and}
  \bibinfo{author}{\bibfnamefont{A.}~\bibnamefont{Komnik}},
  \bibinfo{journal}{Phys. Rev. B} \textbf{\bibinfo{volume}{83}},
  \bibinfo{pages}{075107} (\bibinfo{year}{2011}).

\bibitem[{\citenamefont{Cohen et~al.}(2013)\citenamefont{Cohen, Gull, Reichman,
  Millis, and Rabani}}]{Cohen2013}
\bibinfo{author}{\bibfnamefont{G.}~\bibnamefont{Cohen}},
  \bibinfo{author}{\bibfnamefont{E.}~\bibnamefont{Gull}},
  \bibinfo{author}{\bibfnamefont{D.~R.} \bibnamefont{Reichman}},
  \bibinfo{author}{\bibfnamefont{A.~J.} \bibnamefont{Millis}},
  \bibnamefont{and} \bibinfo{author}{\bibfnamefont{E.}~\bibnamefont{Rabani}},
  \bibinfo{journal}{Phys. Rev. B} \textbf{\bibinfo{volume}{87}},
  \bibinfo{pages}{195108} (\bibinfo{year}{2013}).

\bibitem[{\citenamefont{Jin et~al.}(2008)\citenamefont{Jin, Zheng, and
  Yan}}]{Jin2008}
\bibinfo{author}{\bibfnamefont{J.}~\bibnamefont{Jin}},
  \bibinfo{author}{\bibfnamefont{X.}~\bibnamefont{Zheng}}, \bibnamefont{and}
  \bibinfo{author}{\bibfnamefont{Y.}~\bibnamefont{Yan}}, \bibinfo{journal}{J.
  Chem. Phys.} \textbf{\bibinfo{volume}{128}}, \bibinfo{pages}{234703}
  (\bibinfo{year}{2008}).

\bibitem[{\citenamefont{Zheng et~al.}(2009)\citenamefont{Zheng, Jin, Welack,
  Luo, and Yan}}]{Zheng2009}
\bibinfo{author}{\bibfnamefont{X.}~\bibnamefont{Zheng}},
  \bibinfo{author}{\bibfnamefont{J.}~\bibnamefont{Jin}},
  \bibinfo{author}{\bibfnamefont{S.}~\bibnamefont{Welack}},
  \bibinfo{author}{\bibfnamefont{M.}~\bibnamefont{Luo}}, \bibnamefont{and}
  \bibinfo{author}{\bibfnamefont{Y.}~\bibnamefont{Yan}}, \bibinfo{journal}{J.
  Chem. Phys.} \textbf{\bibinfo{volume}{130}}, \bibinfo{pages}{164708}
  (\bibinfo{year}{2009}).

\bibitem[{\citenamefont{H{\"{a}}rtle et~al.}(2013)\citenamefont{H{\"{a}}rtle,
  Cohen, Reichman, and Millis}}]{Hartle2013}
\bibinfo{author}{\bibfnamefont{R.}~\bibnamefont{H{\"{a}}rtle}},
  \bibinfo{author}{\bibfnamefont{G.}~\bibnamefont{Cohen}},
  \bibinfo{author}{\bibfnamefont{D.~R.} \bibnamefont{Reichman}},
  \bibnamefont{and} \bibinfo{author}{\bibfnamefont{A.~J.}
  \bibnamefont{Millis}}, \bibinfo{journal}{Phys. Rev. B}
  \textbf{\bibinfo{volume}{88}}, \bibinfo{pages}{235426} (\bibinfo{year}{2013}).

\bibitem[{\citenamefont{H\"artle et~al.}(2015)\citenamefont{H\"artle, Cohen,
  Reichman, and Millis}}]{Hartle2015}
\bibinfo{author}{\bibfnamefont{R.}~\bibnamefont{H\"artle}},
  \bibinfo{author}{\bibfnamefont{G.}~\bibnamefont{Cohen}},
  \bibinfo{author}{\bibfnamefont{D.~R.} \bibnamefont{Reichman}},
  \bibnamefont{and} \bibinfo{author}{\bibfnamefont{A.~J.}
  \bibnamefont{Millis}}, \bibinfo{journal}{Phys. Rev. B}
  \textbf{\bibinfo{volume}{92}}, \bibinfo{pages}{085430}
  (\bibinfo{year}{2015}).
  
  \bibitem[{\citenamefont{Cheng et~al.}(2015)}]{Cheng2015}
\bibinfo{author}{\bibfnamefont{Y.}~\bibnamefont{Cheng}},
  \bibinfo{author}{\bibfnamefont{W.}~\bibnamefont{Hou}},
  \bibinfo{author}{\bibfnamefont{Y.}~\bibnamefont{Wang}},
  \bibinfo{author}{\bibfnamefont{Z.}~\bibnamefont{Li}},
  \bibinfo{author}{\bibfnamefont{J.}~\bibnamefont{Wei}}, \bibnamefont{and}
  \bibinfo{author}{\bibfnamefont{Y.}~\bibnamefont{Yan}}, \bibinfo{journal}{New J. Phys.} \textbf{\bibinfo{volume}{17}}, \bibinfo{pages}{033009}
  (\bibinfo{year}{2015}).
  
  \bibitem[{\citenamefont{Wenderoth et~al.}(2016)}]{Wenderoth2016}
\bibinfo{author}{\bibfnamefont{S.}~\bibnamefont{Wenderoth}},
  \bibinfo{author}{\bibfnamefont{J.}~\bibnamefont{B\"atge}},
  \bibnamefont{and} \bibinfo{author}{\bibfnamefont{R.}~\bibnamefont{H\"artle}}, 
  \bibinfo{journal}{Phys. Rev. B}
  \textbf{\bibinfo{volume}{94}}, \bibinfo{pages}{121303R}
  (\bibinfo{year}{2016}).

\bibitem[{\citenamefont{Tanimura and Kubo}(1989)}]{Tanimura1989}
\bibinfo{author}{\bibfnamefont{Y.}~\bibnamefont{Tanimura}} \bibnamefont{and}
  \bibinfo{author}{\bibfnamefont{R.}~\bibnamefont{Kubo}}, \bibinfo{journal}{J.
  Phys. Soc. Jpn.} \textbf{\bibinfo{volume}{58}}, \bibinfo{pages}{101}
  (\bibinfo{year}{1989}).

\bibitem[{\citenamefont{Tanimura}(1990)}]{Tanimura1990}
\bibinfo{author}{\bibfnamefont{Y.}~\bibnamefont{Tanimura}},
  \bibinfo{journal}{Phys. Rev. A} \textbf{\bibinfo{volume}{41}},
  \bibinfo{pages}{6676} (\bibinfo{year}{1990}).

\bibitem[{\citenamefont{Tanimura}(2006)}]{Tanimura2006}
\bibinfo{author}{\bibfnamefont{Y.}~\bibnamefont{Tanimura}},
  \bibinfo{journal}{J. Phys. Soc. Jpn.} \textbf{\bibinfo{volume}{75}},
  \bibinfo{pages}{082001} (\bibinfo{year}{2006}).

\bibitem[{\citenamefont{Welack et~al.}(2006)\citenamefont{Welack, Schreiber,
  and Kleinekath\"ofer}}]{Welack2006}
\bibinfo{author}{\bibfnamefont{S.}~\bibnamefont{Welack}},
  \bibinfo{author}{\bibfnamefont{M.}~\bibnamefont{Schreiber}},
  \bibnamefont{and}
  \bibinfo{author}{\bibfnamefont{U.}~\bibnamefont{Kleinekath\"ofer}},
  \bibinfo{journal}{J. Chem. Phys.} \textbf{\bibinfo{volume}{124}},
  \bibinfo{pages}{044712} (\bibinfo{year}{2006}).

\bibitem[{\citenamefont{Elste et~al.}(2011{\natexlab{b}})\citenamefont{Elste,
  Reichman, and Millis}}]{Elste2011}
\bibinfo{author}{\bibfnamefont{F.}~\bibnamefont{Elste}},
  \bibinfo{author}{\bibfnamefont{D.~R.}~\bibnamefont{Reichman}}, \bibnamefont{and}
  \bibinfo{author}{\bibfnamefont{A.~J.}~\bibnamefont{Millis}},
  \bibinfo{journal}{Phys. Rev. B} \textbf{\bibinfo{volume}{83}},
  \bibinfo{pages}{245405} (\bibinfo{year}{2011}{\natexlab{b}}).

\bibitem[{\citenamefont{Elste et~al.}(2010)\citenamefont{Elste, Reichman, and
  Millis}}]{Elste2010}
\bibinfo{author}{\bibfnamefont{F.}~\bibnamefont{Elste}},
  \bibinfo{author}{\bibfnamefont{D.~R.} \bibnamefont{Reichman}},
  \bibnamefont{and} \bibinfo{author}{\bibfnamefont{A.~J.}
  \bibnamefont{Millis}}, \bibinfo{journal}{Phys. Rev. B}
  \textbf{\bibinfo{volume}{81}}, \bibinfo{pages}{205413} (\bibinfo{year}{2010}).

\bibitem[{\citenamefont{Jang et~al.}(2002)\citenamefont{Jang, Cao, and
  Silbey}}]{Jang2002}
\bibinfo{author}{\bibfnamefont{S.}~\bibnamefont{Jang}},
  \bibinfo{author}{\bibfnamefont{J.}~\bibnamefont{Cao}}, \bibnamefont{and}
  \bibinfo{author}{\bibfnamefont{R.~J.} \bibnamefont{Silbey}},
  \bibinfo{journal}{J. Chem. Phys.} \textbf{\bibinfo{volume}{116}},
  \bibinfo{pages}{2705} (\bibinfo{year}{2002}).

\bibitem[{\citenamefont{Gogolin et~al.}(2004)\citenamefont{Gogolin, Nersesyan,
  and Tsvelik}}]{gogolin2004bosonization}
\bibinfo{author}{\bibfnamefont{A.~O.} \bibnamefont{Gogolin}},
  \bibinfo{author}{\bibfnamefont{A.~A.} \bibnamefont{Nersesyan}},
  \bibnamefont{and} \bibinfo{author}{\bibfnamefont{A.~M.}
  \bibnamefont{Tsvelik}}, \emph{\bibinfo{title}{{Bosonization and Strongly
  Correlated Systems}}} (\bibinfo{publisher}{Cambridge University Press},
  \bibinfo{year}{2004}).



\end{thebibliography}
\end{document}